\documentclass[11pt]{article}
\usepackage[utf8]{inputenc}
\usepackage{fullpage}
\usepackage[T1]{fontenc}
\usepackage{amsmath,amssymb,amsthm}
\usepackage{graphicx,wrapfig, float}
\usepackage{tikz-cd}
\usetikzlibrary{calc}
\usepackage{enumitem,multirow}
\usepackage{physics,relsize,dsfont,cancel,slashed,xfrac}
\usepackage{etoolbox}
\usepackage{import,booktabs,tabularx}
\usepackage{makecell}
\usepackage[pdfpagelabels=false]{hyperref}	
\hypersetup{colorlinks=true,linkcolor=blue, citecolor=purple,filecolor=blue,urlcolor=purple}
\usepackage[titles]{tocloft}
\usepackage[backend=biber, style=phys, biblabel=brackets, articletitle=false, eprint=true]{biblatex}
\usepackage{orcidlink}

\let\oldsqrt\sqrt
\def\sqrt{\mathpalette\DHLhksqrt}
\def\DHLhksqrt#1#2{%
\setbox0=\hbox{$#1\oldsqrt{#2\,}$}\dimen0=\ht0
\advance\dimen0-0.2\ht0
\setbox2=\hbox{\vrule height\ht0 depth -\dimen0}%
{\box0\lower0.4pt\box2}}

\newcommand{\id}{\mathds{1}}
\newcommand{\T}{\scalebox{0.6}{\text{T}}}

\newcommand{\Acal}{\mathcal{A}}

\newcommand{\Gcal}{\mathcal{G}}
\newcommand{\Scal}{\mathcal{S}}

\newcommand{\Ncal}{\mathcal{N}}

\newcommand{\Ccal}{\mathcal{C}}

\newcommand{\dint}{\!\int \! \text{d}^4}

\newcommand{\mI}{m_I^2}

\title{On the Uniqueness of Ghost-Free Multi-Gravity - II:\\
Constraining antisymmetrised multi spin-2 interactions}

\author{
J.~Flinckman\orcidlink{0009-0004-4545-3123},
\quad
S.~F.~Hassan\orcidlink{0000-0003-3910-431X}\\[1ex]
\textit{Department of Physics \& The Oskar Klein Centre,}\\
\textit{Stockholm University, AlbaNova University Centre, SE-106 91 Stockholm, Sweden}\\[1ex]
Email: \href{mailto:joakim.flinckman@fysik.su.se}{\texttt{joakim.flinckman@fysik.su.se}}, \href{mailto:fawad@fysik.su.se}{\texttt{fawad@fysik.su.se}}
}

\date{}

\begin{document}

\maketitle

\begin{abstract} 
So far, only a single theory of multiple spin-2 fields is known that features genuine multi-field interactions while remaining free of Boulware–Deser-type ghost instabilities. In this paper we show that this is the most general ghost-free multi spin-2 interaction type possible. We start with the general class of mutlivielbein interactions containing antisymmetrised products of vielbeins, considered earlier by Hinterbichler and Rosen. We formulate a necessary condition for these theories to be ghost-free. For two vielbeins the theory parameters remain unrestricted, reproducing the ghost-free bimetric theory. But for more than two vielbeins with genuine multi-field interactions, we show that the couplings are restricted precisely to yield the known ghost-free mutlivielbein theory, thus establishing its uniqueness. We also show that more general interactions, constructed using the ghost-free bimetric and mutlivielbein potentials as building blocks, satisfy the necessary ghost-free conditions provided the associated interaction graphs have a tree structure.
\end{abstract}

\tableofcontents

\section{Introduction}

Theories of spin-2 fields interacting with gravity are of interest both from observational and theoretical reasons. Since such fields couple directly to gravity, they are natural candidates for describing dark matter and dark energy, which so far have only been observed via their gravitational effects. Theoretically, such theories generically suffer from ghost instabilities and must rely on novel non-minimal gravitational couplings to be consistent. Such consistency requirements uniquely determine the allowed interactions. Furthermore, the unavoidable non-linear nature of the interactions makes the analysis somewhat challenging.\footnote{Here we restrict to self-contained classical theories with finite field content and hence do not consider, for example, compactifications from higher dimensions.} 

The Einstein–Hilbert action provides consistent dynamics for a massless spin-2 field, non-linearly represented by a metric tensor $g_{\mu\nu}$. This can be easily seen in a 3+1 decomposition in terms of 6 components of the spatial metric $g_{ij}$ which have dynamical equations and the 4 lapse and shift components $N, N^i$ which are non-dynamical (appearing without time derivatives). 

Then as a result of constraints and gauge symmetries, only 2 components of $g_{ij}$ propagate freely, corresponding to the 2 helicities of a massless spin-2 field, as seen for example in gravitational waves. Theories containing more spin-2 fields do not automatically come with the extra gauge symmetries and associated constraints needed to remove the additional degrees of freedom. Hence, reducing the field content from the 10 components of a non-linear spin-2 field down to the 2 helicities of a massless field or the 5 helicities of a massive field becomes non-trivial. Using Einstein–Hilbert kinetic terms allows us to start with 6 dynamical fields for the additional spin-2 fields (instead of 10), but one of these field components is always a ghost mode and appears with a negative kinetic energy term. It is then important that the theory has enough constraints to eliminate this ghost mode as well as all the other non-propagating field components. This requirement puts severe restrictions on the allowed interactions.

So far, there are only two known examples of such theories: the ghost-free bimetric theory \cite{Hassan:2011zd,Hassan:2011ea,Hassan:2018mbl,Hassan:2017ugh}, with two spin-2 fields, and a specific multivielbein theory \cite{Hassan:2018mcw,Flinckman:2024zpb,Flinckman:2025bje} with $\Ncal$ spin-2 fields which is also argued to be ghost-free. Our aim in this work is to investigate the uniqueness of such interactions and their possible generalisations.

For two spin-2 fields, bimetric theory can be regarded as describing the interactions of a spin-2 field $f_{\mu\nu}$ with the gravitational metric $g_{\mu\nu}$. The theory propagates a massless spin-2 mode and a massive spin-2 mode which are combinations of the two metrics. Such a theory was considered earlier in \cite{Isham:1971gm,Chamseddine:1978yu,Damour:2002ws} but was recognized to suffer from a ghost instability due to its similarity with the Boulware–Deser ghost in massive gravity \cite{Boulware:1972yco,Boulware:1972zf}. Much of the subsequent work was in the context of constructing a ghost-free theory of massive gravity which requires the existence of a fixed non-dynamical reference metric, besides the gravitational metric, see e.g.\ \cite{Arkani-Hamed:2002bjr, Creminelli:2005qk}, culminating in ghost-free versions of massive gravity with flat reference metric $\eta_{\mu\nu}$ \cite{deRham:2010ik,deRham:2010kj}. The absence of the ghost in this model, and its extension to a generic reference metric, was established in \cite{Hassan:2011hr,Hassan:2011tf,Hassan:2011ea,Hassan:2012qv,Comelli:2013txa} and several other works. Bimetric theory is an extension of this framework to a theory of two dynamical, interacting spin-2 fields. The theory has interesting theoretical features \cite{Hassan:2017ugh,Kocic:2018yvr} besides being ghost-free, and its phenomenological applications have been investigated in many works, see e.g.\ \cite{vonStrauss:2011mq,Akrami:2012vf,Enander:2015kda,DeFelice:2013nba}.

Having established a theory of two spin-2 fields, the next obvious question is its extension to multiple spin-2 fields. A simple such extension consists of many spin-2 fields $g_{I\mu\nu}$ that only interact pairwise via the bimetric potential. The basic ghost-free cases are interactions of the form $V(g_1,g_2) + V(g_2,g_3) + V(g_3,g_4)+\cdots$ and $V(g_1,g_2) + V(g_1,g_3) + V(g_1,g_4)+\cdots$ \cite{Khosravi:2011zi,Molaee:2018brt, Dokhani:2020jxb}.  However, it is more interesting to look for genuine multi-spin-2 interactions. One such theory was proposed in \cite{Hinterbichler:2012cn}, where the interaction involves an antisymmetrised product of vielbeins, hence each interaction term involves a product of at most $D$ different vielbeins in $D$ dimensions. In 4 spacetime dimensions, and for $\Ncal$ vielbeins $e_I$ ($I=1,\cdots,\Ncal$), this theory has coupling parameters $\beta^{IJKL}$, symmetric in all indices. Arguments were provided that this theory contained the correct constraint structure to be ghost-free. However, a loophole in the arguments was found by considering special interactions allowed by this construction and showing that the full set of constraints did not have the correct form to eliminate the ghosts \cite{Nomura:2012xr,Scargill:2014wya,Afshar:2014dta}, leading to the possibility that ghost-free spin-2 interactions may not exist beyond bimetric interactions \cite{deRham:2015cha}.

More recently, a theory with multiple spin-2 interactions was constructed which circumvents the no-go arguments found in the earlier context. This theory has a simple multivielbein interaction potential given in terms of the determinant of a sum of the vielbeins $e_I$ with couplings $\beta^I$ as $\det(\beta^1 e_1+\cdots+\beta^\Ncal e_\Ncal)$. It admits a healthy (non-tachyonic) spectrum of one massless and $\Ncal-1$ massive spin-2 modes \cite{Flinckman:2024zpb} and has been argued to possess the full set of constraints needed to eliminate the ghost modes \cite{Flinckman:2025bje}. Compared to the earlier model in \cite{Hinterbichler:2012cn}, the ghost-free theory corresponds to choosing the couplings as $\beta^{IJKL}=\beta^I\beta^J\beta^K\beta^L$.

In this work, and an upcoming companion paper \cite{Flinckman:2026xx}, we investigate the possibility of finding more general ghost-free multi-spin-2 interactions. In particular, for $\Ncal=2$ it is easy to see that the multivielbein potential $\det(\beta^1e_1 + \beta^2 e_2)$ recovers only a subset of the known bimetric interactions, so clearly in this case the theory admits a generalisation. The situation is more complicated for $\Ncal>2$. To simplify the analysis, we set out to formulate a necessary condition for the absence of ghosts by considering a well-motivated restricted configuration of vielbeins by eliminating the shift and the Lorentz boost parameters. We search for possible ghost-free models within the general class of antisymmetrised vielbein interactions of \cite{Hinterbichler:2012cn}. While here we take the Hinterbichler–Rosen potential as our starting point, the upcoming companion paper investigates its uniqueness and possible generalisations.

\subsection{Results and summary of the paper}
\label{summary}

Since the rest of the paper is somewhat technical, here we present a detailed summary of the reasoning and results. 

We consider the class of multivielbein models introduced in \cite{Hinterbichler:2012cn} for more than two vielbeins and precisely formulate the obstructions that prevent them from being ghost-free. We then look for possible theories within this class that circumvent the generic obstructions and, thus, could potentially be ghost-free. We show that the multivielbein theory constructed in \cite{Hassan:2018mcw} is the only such possibility, along with its simple extensions.

The analysis follows the reasoning: In terms of vielbein 1-forms $e^A_I=e^A_{I\mu}\mathrm{d}x^\mu$ (for $I=1,\cdots,\Ncal$), the interactions introduced in  \cite{Hinterbichler:2012cn} are $\beta^{IJKL}\epsilon_{ABCD}\,e^A_{~I}\wedge e^B_{~J}\wedge e^C_{~K}\wedge e^D_{~L}$, with sums over repeated indices. The $\beta^{IJKL}$ are totally symmetric couplings that we wish to constrain. We first identify an {\it irreducible} set of the couplings to exclude known pairwise bimetric interactions as well as models that contain clusters of interacting vielbeins with only one vielbein shared between two different clusters (these are analysed separately later). We then set out to restrict the irreducible $\beta^{IJKL}$ by analysing the constraints. 

To isolate the relevant ghost removing constraints, we use the fact that in a 3+1 decomposition every vielbein $e^A_{~\mu}$ can be decomposed into 6 Lorentz parameter fields (for 3 boosts and 3 spatial rotations), 4 lapse and shift variables $N, N_i$ and finally the 6 components of the spatial metric $g_{ij}$ or, equivalently, a gauge fixed spatial vielbein $\bar{e}^a_{~i}$. Einstein–Hilbert terms in the action give dynamics only to the 6 components $\bar{e}^a_{~i}$ of each vielbein. These include one ghost field per vielbein along with 5 non-ghostly dynamical fields. The remaining fields have non-dynamical equations of motion. For the theory to be ghost-free, these non-dynamical equations must contain constraints that can eliminate the ghost fields contained in the spatial vielbeins $\bar{e}^a_{~i}$. From general symmetry arguments it follows that the ghost modes in the spatial metrics must be eliminated, if at all, by the equations of motion for the lapses, both being scalars under 3 dimensional spatial rotations and spatial diffeomorphisms. Hence to find a necessary condition for the absence of ghosts, it is enough to consider vielbein configurations without the shift and boost variables, considerably simplifying the analysis. We will still retain the spatial rotation fields in the vielbeins since they are a source of the obstructions. We then analyse the non-dynamical equations in this setup and find the obstructions that prevent the removal of ghosts for generic couplings $\beta^{IJKL}$.

We then consider the conditions under which the equations for the lapses and Lorentz rotation fields have the correct form to be able to eliminate the ghost modes. Using a theorem from tensor analysis, we classify the irreducible couplings $\beta^{IJKL}$ based on their {\it symmetric tensor rank $R$} and show that the ghost removal conditions require $R=1$. This is precisely the case when $\beta^{IJKL}=\beta^I\beta^J\beta^K\beta^L$, corresponding to the ghost-free multivielbein model found in \cite{Hassan:2018mcw}, showing that for $\Ncal>2$, this is the unique ghost-free irreducible interaction. The arguments for the absence of ghost in this model, beyond the necessary conditions found  here, are given in \cite{Hassan:2018mcw,Flinckman:2025bje}.  

Finally we describe an algorithm for building {\it reducible} ghost-free interactions by combining the basic ghost-free bimetric and multimetric interactions. Several examples of reducible interactions are then discussed in detail in the appendix. 

The rest of the paper is organized as follows: in Section \ref{sec:known_ghost_free} we review the known ghost-free bimetric and multivielbein interactions and their simple extensions, along with the general antisymmetrised multivielbein potential that is analysed later. In Section \ref{sec:restric} we characterise the irreducible interactions and setup the framework for the analysis of the relevant constraints. We find the necessary condition for the irreducible interactions to be ghost-free. In Section \ref{sec:generic_trees}, we provide an algorithm for building reducible ghost-free interactions starting with irreducible ones. Explicit examples of this are relegated to appendix \ref{app:examples}. Section \ref{sec:conclusion} contains a brief conclusion and discussion of results.   

\section{Known ghost-free multimetric theories}
\label{sec:known_ghost_free}
In this section we review the known ghost-free theories of spin-2 fields interacting with gravity, restricting to non-derivative interactions between different spin-2 fields. 
These fall in two classes:
\begin{enumerate}
    \item Theories with bimetric type interactions: These include the ghost-free bimetric theory and its direct extensions to multiple spin-2 fields with only pairwise bimetric couplings.
    \item Multivielbein theories based on a determinant potential: These describe interactions of more than two spin-2 fields with genuine multi-field interactions encoded in a specific determinant interaction.
\end{enumerate}
We will then address the question of uniqueness and possible generalisations of the ghost-free multivielbein interactions. 

\subsection{A review of bimetric theory}
\label{sec:rev_bimetric}
Ghost-free bimetric theory is a theory of two spin-2 fields, usually denoted by $g_{\mu\nu}$ and $f_{\mu\nu}$, each governed by an Einstein–Hilbert action, and interacting via a non-derivative potential $V_{\mathrm{bi}}$ which has been shown to be free of non-linear ghost instabilities 
\cite{Hassan:2011zd,Hassan:2011ea,Hassan:2018mbl},
\begin{align}
\label{Bim}
    \Scal_{\mathrm{bi}}=\dint x\left[ m_g^2\sqrt{-g}R^{(g)}+m_f^2 \sqrt{-f}R^{(f)} -V_{\text{bi}}(g,f) \right]+\Scal_{\mathrm{matter}}\,,    
\end{align}
where $m_g$ and $m_f$ are the two Planck masses. The metrics enter the potential only through the combination $S^\mu_{\,\nu}=(\sqrt{g^{-1}f}\,)^\mu_{\,\nu}$ which is the the square root of the matrix $g^{\mu\lambda}f_{\lambda\nu}$, so that $S^2=g^{-1}f$. Then $V_{\text{bi}}(g,f)$ is given in terms of the elementary symmetric polynomials $P_n(S)$ of the matrix $S$ with coefficients $\alpha_n$ as,
\begin{align}
\label{bi_pot}
    V_{\text{bi}}(g,f) = 2m^4\sqrt{-g}\sum_{n=0}^4 \alpha_n P_n(S).
\end{align}
The $P_n(S)$ are explicitly given by (with $[S]=\Tr(S)$, etc.),
\begin{equation}
\label{el_pol}
\begin{aligned}
    P_0(S)&=1,\qquad P_1(S)=[S],\qquad P_2(S)= \tfrac{1}{2}\left([S]^2-[S^2]\right)\\
    P_3(S)&=\tfrac{1}{6}\left([S]^3-3[S][S^2]+2[S^3]\right),\qquad P_4(S)= \det(S)\,.
\end{aligned}
\end{equation}
Note that $\alpha_0$ and $\alpha_4$ enter as cosmological terms for the two metrics (though all $\alpha_n$ will contribute to the observed cosmological constant). Each metric can couple to its own matter sector as $\Scal_{\mathrm{matter}}=
\Scal_1(g,\psi_g)+\Scal_2(f, \psi_f)$ with no direct interactions between the two matter sectors. On setting $f_{\mu\nu }=\eta_{\mu\nu}$ by hand and ignoring its equation of motion, the above model reduces to the massive gravity model of \cite{deRham:2010ik,deRham:2010kj} in which the absence of ghost was proven in \cite{Hassan:2011hr,Hassan:2011tf,Hassan:2011ea}. The uniqueness of the dRGT massive gravity potential, and hence of the non-derivative bimetric potential above, follows, for example, from the analysis of \cite{Comelli:2013txa, Molaee:2018brt}.

Alternatively, the potential $V_{\text{bi}}(g,f)$ can be expressed as an antisymmetric product of $S^\mu_{~\nu}$,
\begin{align}
V_{\text{bi}}(g,f)= 2m^4\sqrt{-g}
\sum_{n=0}^4\frac{\alpha_n}{n!(4-n)!}
\epsilon_{\mu_1\cdots\mu_n\lambda_{n+1}\cdots\lambda_4}    
\epsilon^{\nu_1\cdots\nu_n\lambda_{n+1}\cdots\lambda_4}
S^{\mu_1}_{~\nu_1}\cdots S^{\mu_n}_{~\nu_n} .
\end{align}
Introducing vielbeins $e^A_{1\,\mu}$ and $e^A_{2\,\mu}$ such that, in matrix notation, $g=e^{\T}_1\eta e^{\,}_1$ and $f=e^{\T}_2\eta e^{\,}_2$, one can easily verify that if the vielbeins satisfy a symmetrisation condition $e^A_{1\,[\mu}\eta_{AB}e^B_{2\,\nu]}=0$ or, in matrix notation, 
\begin{align}
\label{bim-symmetrization}
    e_1^{\T}\eta e^{\,}_2 = e_2^{\T}\eta e_1^{\,}\,,    
\end{align}
then, the square-root matrix evaluates to $S^\mu_{~\nu}=e^\mu_{1\,A}e^A_{2\,\nu}$, where $e^\mu_{1\,A}= (e_1^{-1})^\mu_{~A}$. This motivates expressing the bimetric potential in term of the vielbeins as \cite{Hinterbichler:2012cn} (see also \cite{Zumino:1970tu}), 
\begin{align}
\label{bi_wedge}
V_{\text{bi}}(e_1,e_2) = 2m^4
\sum_{n=0}^4\frac{\alpha_n}{n!(4-n)!}
\epsilon_{A_1\cdots A_n A_{n+1}\cdots A_4}
\epsilon^{\mu_1\cdots\mu_n\mu_{n+1}\cdots\mu_4}\,    
e^{A_1}_{2\,\mu_1}\cdots e^{A_n}_{2\,\mu_n} e^{A_{n+1}}_{1\,\mu_{n+1}}\cdots e^{A_4}_{1\,\mu_4},
\end{align}
where, $e_1$ and $e_2$ are now unconstrained vielbeins. The symmetrization condition \eqref{bim-symmetrization} which establishes the equivalence to the metric formulation then emerges as an equation of motion in the vielbein theory as follows:

The equation of motion for the vielbein $e_1$ has the form,
\begin{align}
    R^{(g)}_{\mu\nu}-\tfrac{1}{2}g_{\mu\nu}R^{(g)} + V^{(g)}_{\mu\nu}=\tfrac{1}{2m_g^2}T^{(g)}_{\mu \nu}.
\end{align}
Since the first two and last terms are explicitly symmetric, it follows that the antisymmetric part of the remaining term must vanish, $V^{(g)}_{[\mu\nu]}=0$. These are the equations of motion for the Lorentz parameters contained in the vielbein $e^A_{1\,\mu}$ \cite{Hinterbichler:2012cn,Hassan:2012wt,Flinckman:2024zpb}, therefore known as the Lorentz constraints,  
and imply the symmetrisation condition \eqref{bim-symmetrization}.\footnote{The condition $V^{(g)}_{[\mu\nu]}=0$ may also admit other solutions besides \eqref{bim-symmetrization} but these are not matrix solutions and are not consistent with Lorentz and general coordinate transformations \cite{Hassan:2017ugh}.} The equation of motion for the vielbein $e_2$ leads to exactly the same conclusion.

In the bimetric case, the metric and vielbein formulations are equivalent: First, note that a well defined bimetric theory exists only if the square-root matrix $S=\sqrt{g^{-1}f}$ can be uniquely specified as a real (1,1) tensor. Such an $S$ exists if and only if (i) the null cones of the metrics $g_{\mu\nu}$ and $f_{\mu\nu}$ have a non-vanishing intersection such that they admit common spacelike hypersurfaces, and (ii) $S$ is chosen as the principle square root \cite{Hassan:2017ugh}. Then, starting from generic vielbeins for the metrics $g_{\mu\nu}$ and $f_{\mu\nu}$ one can always find local Lorentz transformations that lead to vielbeins $e_1$ and $e_2$ satisfying \eqref{bim-symmetrization}. Conversely, given two real vielbeins that satisfy \eqref{bim-symmetrization}, a real square-root matrix $S=\sqrt{g^{-1}f}$ exists \cite{Hassan:2017ugh, Deffayet:2012zc}.

\subsection{Multimetric theories with pairwise bimetric interactions}
\label{sec:bi_pairwise}
The main feature of the bimetric theory \eqref{Bim} is that the specific combination of the Einstein–Hilbert terms and the non-derivative potential $V_{\text{bi}}(g,f)$ is ghost-free. It can be seen that the basic property of \eqref{Bim}, which is required for the absence of ghost, remains preserved in certain simple extensions to multiple spin-2 fields as long as the metrics interact only pairwise via the bimetric potential \cite{Khosravi:2011zi,,Molaee:2018brt, Dokhani:2020jxb}. There are two basic types of such extensions, 
\begin{align}
\label{eq:pairwise}
    V_{\text{centre}}= \sum_{I=2}^{\mathcal{N}}V_{\text{bi}}(g_1,g_{I}), && V_{\text{chain}}= \sum_{I=1}^{\mathcal{N}-1}V_{\text{bi}}(g_I,g_{I+1}),
\end{align}
which can be combined as long as the interaction graph, with metrics $g_I$ as vertices $\bullet$ and bimetric potentials $V_{\text{bi}}(g_I,g_J)$ as edges - connecting the two vertices $\bullet$–$\bullet$, forms a tree, i.e.\ a connected  graph on $\Ncal$ vertices with $\Ncal-1$ edges and no cycles. The existence of a cycle in such a setup reintroduces ghosts \cite{Nomura:2012xr,Scargill:2014wya,Afshar:2014dta}, see Appendix \ref{sec:bi_cycle}.

Such pairwise multimetric theories can also be formulated in terms of vielbeins, with the bi-vielbein potentials $V_{\text{bi}}(e_I,e_{J})$ \eqref{bi_wedge} replacing the bimetric ones in \eqref{eq:pairwise}. Then, the tree structure guarantees that the Lorentz constraints always admit pairwise symmetrisation, i.e., if $e_I$ and $e_J$ are connected by an edge in the tree, the Lorentz constraints will impose,
\begin{align}
\label{pairwise_sym}
    e_I^{\T}\eta e_J^{\,} = e_J^{\T}\eta e_I^{\,}.
\end{align}
This in turn ensures equivalence to the metric formulation. However, in the presence of a cycle, the Lorentz constraints do not imply symmetrizations of individual vielbein pairs and equivalence to the metric formulation is lost.

\subsection{A Ghost-free Multivielbein theory}

A ghost-free theory of multiple interacting spin-2 fields with genuine multi-field couplings beyond pairwise bimetric interactions was first formulated in \cite{Hassan:2018mcw}, and argued to be ghost-free in \cite{Flinckman:2025bje}. While each spin-2 field $g^I_{\mu \nu}$, $I=1,\ldots, \Ncal$, is governed by an Einstein–Hilbert action, the interaction is formulated in terms of the associated vielbeins $e^A_{I\,\mu}$,
\begin{align}
\label{det_action}
    \Scal = \dint x\left[\sum_{I=1}^\Ncal \mI \sqrt{-g^{\,}_{I}}(R^{I}-2\Lambda_I)   - V_{\mathrm{det}}(e_1,\cdots, e_\Ncal)\right] +\Scal_{\mathrm{matter}}.
\end{align}
Here, $m_I$ are Planck masses and $\Lambda_I$ cosmological constants. The potential $V_{\mathrm{det}}$ is  given by the determinant of a sum of vielbeins,
\begin{align}
\label{det_coupling}
    V_{\mathrm{det}} &= 2m^4\det(u), && u =\sum_{I=1}^\Ncal\beta^I e_I.
\end{align}
For non-trivial interactions to exist, we require that $\det(u)\neq 0$. As in bimetric theory, in $\Scal_{\mathrm{matter}}$, each vielbein can couple to its own matter sector $\Scal_I(e_I, \psi_I)$, but a matter field cannot couple to more than one vielbein, nor interact directly with fields from another sector.

On expanding the determinant, the interaction can be written as an antisymmetric product of vielbeins,
\begin{align}
    V_{\mathrm{det}} = \frac{2m^4}{4!}\!\!\!\!\sum_{IJKL=1}^\Ncal\!\!\!\!\beta^I \beta^J\beta^K \beta^L \epsilon_{ABCD}\epsilon^{\alpha \beta \gamma \delta}e^A_{I\, \alpha}e^B_{J\, \beta}e^C_{K\, \gamma}e^D_{L\, \delta},
\end{align}
similar to the bimetric form \eqref{bi_wedge}. However, this theory does not yet have a known formulation entirely in terms of the metrics.\footnote{Even though the reformulation of \cite{Hassan:2012wt} can be used, the result will still contain the Lorentz degrees of freedom explicitly.} From this form, it is evident that the determinant coupling goes beyond the pairwise bimetric interactions, since the potential contains products of more than two distinct vielbeins, e.g. $\beta^1\beta^1\beta^2\beta^3e_1 e_1 e_2 e_3$, which cannot arise from pairwise bimetric interactions. Notably, the expansion contains pairwise bimetric cycles together with other multi-field terms, each of which we will later see is individually inconsistent. With the determinant coefficients their obstructions cancel, and their presence is in fact necessary for ghost-freedom, see Appendix~\ref{app:curing}.

The equations of motion for the vielbein $e_I$ take the form,
\begin{align}
\label{mm_field_Eq}
    R^I_{\mu \nu}-\tfrac{1}{2}g^I_{\mu \nu}R^I +\Lambda_I g^I_{\mu \nu}+V^I_{\mu \nu}= \tfrac{1}{2m_I^2}T^I_{\mu \nu},
\end{align}
where $V^I_{\mu \nu}$ arises from the multivielbein interaction. Since all terms but $V^I_{\mu \nu}$ are manifestly symmetric, one gets the Lorentz constraints $V^{I}_{[\mu \nu]}=0$. It is easy to see \cite{Hassan:2018mcw} that these imply the symmetrisation of each vielbein $e_I$ with the sum $u = \sum_J \beta^J e_J$, 
\begin{align}
\label{eetau=}
    e_I^{\T}\eta u = u^{\T}\eta e_I,
\end{align}
providing a generalisation of the bimetric symmetrisation condition.

Note that, for two vielbeins, $\Ncal=2$, this condition yields precisely \eqref{bim-symmetrization}, and the determinant interaction reproduces a subset of bimetric interactions \eqref{bi_pot},
\begin{align}
    V_{\mathrm{det}} = 2m^4 \det ( \beta_1e_1 + \beta_2e_2)&= 2m^4 \sqrt{-g}\det(\beta_1 \id + \beta_2 e_1^{-1}e_2)\notag\\
    &=2m^4 \sqrt{-g}\sum_{n=0}^4 \beta_1^{4-n}\beta_2^nP_n(S),
\end{align}
with a restricted set of bimetric couplings $\alpha_n = \beta_1^{4-n}\beta_2^n$ for $n=1,2,3$, while $\alpha_0=\beta_1^4+m_1^2\Lambda_1/m^4$ and $\alpha_4=\beta_2^4+m_2^2\Lambda_2/m^4$. Since $\beta_1$ and $\beta_2$ span only a two-parameter subset of the three bimetric couplings $\alpha_1, \alpha_2$ and $\alpha_3$, we find that for $\Ncal=2$, the multivielbein potential \eqref{det_coupling} is not the most general ghost-free interaction possible. Rather, it can be extended to the full class of ghost-free bimetric theories. One might therefore ask whether such ghost-free generalizations also exist for $\Ncal> 2$. We address this issue in the following sections.

\subsection{Multivielbein theories with combined determinant interactions}

Just as the bimetric potential can be used to build larger multi-gravity theories through pairwise couplings arranged in a tree, the determinant interaction \eqref{det_coupling} can be extended by combining multiple determinant sectors. Consider two groups of vielbeins with labels in the sets $\Acal_1$ and $\Acal_2$  sharing exactly one vielbein, say $e_1$, 
(e.g. $\Acal_1=\{1,2,3,4\}$ and $\Acal_2=\{1,5,6,7\}$) with the potential,
\begin{align}
\label{det_sum_potential}
    V = 2m^4\bigg[\det\Big(\sum_{I\in \Acal_1}\beta^I_1 e_I\Big) + \det\Big(\sum_{I\in \Acal_2}\beta^I_2 e_I\Big)\bigg].
\end{align}
Then $e_1$ interacts with all the vielbeins, while the other vielbeins in $\Acal_1$ and $\Acal_2$ interact only directly with among themselves with a determinant interaction. For $e_I \in \Acal_1\setminus\{1\}$, the Lorentz constraint reads,
\begin{align}
    e_I^{\T}\eta u_1 = u_1^{\T}\eta e_I, && u_1=\sum_{I\in \Acal_1} \beta^I_1 e_I,
\end{align}
and similar for $e_I \in \Acal_2\setminus\{1\}$. Then the Lorentz constraint for the common vielbein $e_1$ involves a combination of the two constraints above and is automatically satisfied. This is the crucial feature of this type of extension which preserves the simpler Lorentz constraints of the individual determinent interactions. 

This construction generalises to any number of determinant sectors $\Acal_1,\ldots,\Acal_R$, each carrying its own determinant coupling, provided any two sectors share at most one vielbein and the resulting graph of sectors $\Acal_r$ again forms a tree. The tree structure guarantees that the Lorentz constraints always reduce to,
\begin{align}
\label{mm-symmetrzation}
    e_I^{\T}\eta u^{\,}_{r}=u_{r}^{\T}\eta e^{\,}_I, && u_{r}= \sum_{I\in \Acal_r}\beta^I_r e_I
\end{align}
for a vielbein in $\Acal_r$, something we will elaborate on in Section~\ref{sec:generic_trees}. 

Our aim is to find out if genuine ghost-free multivielbein interactions exist beyond the bimetric and determinant couplings and their simple extensions described above.

\subsection{General antisymmetrised multi spin-2 interaction}

In 2012 Hinterbichler and Rosen presented a multi-gravity potential formulated fully in terms of an antisymmetric product of $\Ncal$ vielbeins \cite{Hinterbichler:2012cn}. Their action reads,
\begin{align}
\label{MM_action}
    \Scal = \dint x\left[\sum_{I=1}^\Ncal \mI \sqrt{-g^{\,}_{I}} R^{}_I - V_{\mathrm{HiRo}}(e_1,\cdots,e_\Ncal)\right] +\mathcal{S}_{\text{matter}},
\end{align}
with a potential of the form,
\begin{align}
\label{pot}
    V_{\mathrm{HiRo}} &= \frac{2m^4}{4!}\!\!\!\!\sum_{IJKL=1}^\Ncal \!\!\!\!\beta^{IJKL}\epsilon_{ABCD}\epsilon^{\alpha \beta \gamma \delta}e^A_{I\, \alpha}e^B_{J\, \beta}e^C_{K\, \gamma}e^D_{L\, \delta},
\end{align}
for generic totally symmetric couplings $\beta^{IJKL}$.

The field equations take the same form as \eqref{mm_field_Eq}, now with $V^I_{\mu \nu}$ obtained from the generalised multivielbein interaction \eqref{pot}. Its symmetry, imposed by the equations of motion, implies $V^I_{[\mu \nu]}=0$ which after some manipulations yields the Lorentz constraints,  
\begin{align}
\label{wedge_LC}
     \Ccal^{AB}_I =\sum_{JKL}\epsilon^{\alpha \beta \gamma \delta} \beta^{IJKL}(e_I^{\T}\eta e_J^{\;})_{\alpha \beta}e^{[A}_{K \; \gamma}e^{B]}_{L \; \delta}=0.
\end{align}
For generic $\beta^{IJKL}$ these do not reduce to a bilinear symmetrisation of the vielbeins in contrast to \eqref{pairwise_sym} and \eqref{mm-symmetrzation}.

The potential \eqref{pot} clearly contains the determinant interaction through the choice $\beta^{IJKL}= \beta^I \beta^J \beta^K \beta^L$, but also reproduces the bimetric interaction in the form \eqref{bi_wedge} for $\Ncal=2$, where,
\begin{align}
\label{bi_beta}
    \alpha_n = \beta^{\overbrace{ \!\scriptstyle1\ldots 1}^{4-n}\overbrace{\scriptstyle 2\ldots 2}^{n}}.
\end{align}
In Section~\ref{sec:generic_trees}, we will also see how \eqref{pot} contains the pairwise bimetric interactions and sums of determinants discussed above. However, for generic $\beta^{IJKL}$ the potential \eqref{pot} is not ghost-free. For instance, generic $\beta^{IJKL}$ allows for pairwise bimetric interactions forming cycles, which are inconsistent \cite{Nomura:2012xr,Afshar:2014dta,Scargill:2014wya,deRham:2015cha}.

In the next section, we analyse the constraint structure of \eqref{MM_action} for general $\beta^{IJKL}$ and derive necessary conditions for the theory to be ghost-free. This will allow us to find restrictions on the allowed $\beta^{IJKL}$ and single out only one consistent genuine multi-gravity interaction.

\section{Restricting the antysimmetrised potential}
\label{sec:restric}

\subsection{Lapse dependent constraints}

The Hinterbichler–Rosen interaction is not ghost-free for arbitrary coupling parameters $\beta^{IJKL}$. The problem arises in the constraint structure, where if the field equations for the non-dynamical variables determine all of them directly, no additional constraints remain to eliminate the ghost modes contained in the spatial vielbeins $\bar{E}^a_{~i}$.

To make this precise, we perform a $3{+}1$ decomposition of each metric,
\begin{align}
    g^I_{\mu \nu}\text{d}x^\mu \text{d}x^\nu = -N^2_I \text{d}t^2 +\gamma^I_{ij}\big(\text{d}x^i+ N^i_I \text{d}t\big)\big(\text{d}x^j+ N^j_I \text{d}t\big),
\end{align}
into lapses $N_I$, shifts $N_I^i$, and spatial metrics $\gamma^I_{ij}$. Additionally, each vielbein can be factorised,
\begin{align}
\label{e=Lebar}
    e^A_{I\, \mu}= L^A_{I\, B}\overline{e}^B_{I\, \mu},
\end{align}
where each $L^A_{I\, B}$ parametrise the 6 Lorentz fields (3 boosts and 3 rotations) and,
\begin{align}
\label{ebar}
    \overline{e}^A_{I\, \mu}= \begin{pmatrix}
        N_I & 0\\
        \bar{E}^b_{I\;j}N^j_I &  \bar{E}^b_{I\;i}
    \end{pmatrix},
\end{align}
depends only on the metric fields and $\bar{E}^a_{I\, i}$ is a spatial vielbein of $\gamma_{ij}^I=(\bar E^{\T}_I \delta \bar E^{\,}_I)_{ij}$.

With this decomposition, the Hinterbichler–Rosen action is linear in the lapses and shifts,
\begin{align}
\label{phase_space_action}
    \Scal = \dint x \sum_I\Big[\pi^i_{I\, a}\dot{\bar{E}}^a_{I\, i} + N_I \Ccal^I + N^i_I \Ccal^I_i \Big],
\end{align}
and the functions $\Ccal^I$ and $\Ccal^I_i$ depend only on the spatial vielbeins $\bar E_I$ and the Lorentz variables $L_I$. The lapse $N_I$ and shift $N_I^i$ therefore appear as Lagrange multipliers, and their equations of motion impose the constraints,
\begin{align}
    \Ccal^I(L,\bar E,\pi) =0, && \Ccal^I_i(L,\bar E,\pi) =0.
\end{align}
For the theory to be ghost-free, it is crucial that the conditions $\Ccal^I=0$ constrain the spatial vielbeins $\bar{E}^a_{~i}$ and their conjugate momenta $\pi^i_{~a}$, which together contain the ghost modes, rather than determine the lapses $N_I$.  However, solving the Lorentz constraints \eqref{wedge_LC}, in particular for the rotations, generically makes $\Ccal^I$ depend on the lapses.\footnote{Note that the boost parameters can always be determined independently of the lapses and shifts by solving $\Ccal^i_I=0$ for them.} The equations $\Ccal^I(L(N))=0$ then determine the lapses $N_I$ instead of providing the additional constraints needed to eliminate the ghost modes.

To see this explicitly, note that only the first column of each vielbein \eqref{e=Lebar} depends on the lapses and shifts,
\begin{align}
\label{e_lapse_lin}
    e^A_{I~0}= N_I L^A_{I\, 0} + N^i_Ie^A_{I\,i}
\end{align}
Since the $\epsilon$-symbol of \eqref{wedge_LC} only picks up one $e^A_{I\, 0}$ at a time, the Lorentz constraints are linear in lapses and shifts,
\begin{align}
\label{lin_LC}
    \Ccal^{AB}_I = \sum_J \Big[N_JF^{AB}_{IJ} + N_J^iG^{AB}_{i\, IJ} \Big]=0,
\end{align}
for some functions $F^{AB}_{IJ}$ and $G^{AB}_{i\, IJ}$ dependent only on the spatial vielbeins $\bar{E}_I$, the boosts and the rotations in $L_I$. Since these equations are linear in the lapses and shifts, the solutions for the Lorentz fields generically depends on $N_I$ and $N^i_I$, so that $L^A_{I\,B}(N,N^i)$.

Once the Lorentz constraints \eqref{lin_LC} have been solved, the equations $\Ccal^I_i(L(N,N^i))=0$ determine the shifts $N^i_I$, and $\Ccal^I(L(N,N^i))=0$ determine the lapses $N_I$.\footnote{In the original paper \cite{Hinterbichler:2012cn}, the authors solve $\mathcal{C}^I_i=0$ for the boosts in terms of the spatial vielbeins (including rotations), their momenta, and the shifts, and the shifts are determined from parts of the Lorentz constraints. Fixing the shifts is not problematic for ghost-freedom; the problem arises only if the lapses are determined, since then the corresponding constraints $\Ccal^I=0$ are lost and cannot eliminate the ghost modes.} At this point, all non-dynamical variables have been determined, leaving no room for the additional constraints needed to eliminate the ghost modes.\footnote{In the Hamiltonian formalism, this is reflected in the fact that preserving the constraints $\Ccal^I=0$, $\Ccal^I_i=0$, and $\Ccal^{AB}_I=0$ in time determines the Lagrange multipliers associated with the primary constraints given by the vanishing conjugate momenta of $N_I$, $N_I^i$, and $L^A_{I\,B}$.} For the interaction \eqref{pot} to be ghost-free, it is therefore \emph{necessary} that $\beta^{IJKL}$ is such that the Lorentz constraints \eqref{lin_LC} can be solved for the rotations independently of the lapses. 

Before we proceed further, recall that the Lorentz constraints reduce to \eqref{pairwise_sym} in the pairwise bimetric case and to \eqref{eetau=} for the multi-determinant couplings. These can equivalently be written,
\begin{align}
    \textbf{Bimetric:} &&[e_I^{\T}\eta e_J^{\;}]_{[\mu \nu]}=0,\\
    \textbf{Determinant:}&&[e_I^{\T}\eta u]_{[\mu \nu]}=0, && u= \sum_I \beta^I e_I.
\end{align}
While these constraints are not entirely lapse-independent, their spatial components $\mu\nu=ij$ are, and these are precisely the components that determine the rotations. Indeed, $[e_I^{\T}\eta e_J^{\;}]_{[ij]}$ and $[e_I^{\T}\eta u]_{[ij]}$ depend only on the spatial components $e^A_{~i}$ of the vielbeins, whereas the lapse enters only through the time components $e^A_{~0}$ \eqref{e_lapse_lin}. It thus follows that, in both the bimetric and determinant interaction the rotations can be solved independent of the lapses. If the lapse independent $\Ccal^I_i=0$ are used to determine the boost parameters, all the Lorentz fields are determined and $\Ccal^I=0$ remains independent of lapses therefore constraining the dynamical fields, eliminating a ghost mode.

We will now use the notion of lapse independent rotations to derive necessary conditions on $\beta^{IJKL}$.

\subsection{Shiftless Ansatz}

To simplify the analysis, we restrict the vielbeins to a block diagonal form,
\begin{align}
\label{mini_ansatz}
    e^A_{I\; \mu} = 
    \begin{pmatrix}
        N_I & 0 \\
        0 & E^a_{I\, i}
    \end{pmatrix}, \qquad E^a_{I\,i}= \Omega^a_{I\, b}\bar{E}^b_{I\,i}
\end{align}
effectively setting all boosts and shifts to zero, and parametrising the rotations in terms of $\Omega_I$. This is sufficient for our purposes: the ghost modes reside in the spatial metrics and transform as scalars under spatial rotations and spatial diffeomorphisms. They can therefore be eliminated, if at all, only by constraints with the same transformation properties, namely the scalar lapse equations. It is thus enough to isolate this sector while retaining the rotations that source the obstruction. If the Lorentz constraints already overconstrain the system in this restricted setting, restoring the additional variables can only exacerbate the problem.

With \eqref{mini_ansatz} the Lorentz constraints \eqref{wedge_LC} reduce to,
\begin{align}
\label{reduced_LC}
    \Ccal^{a}_I &= \sum_{JKL}N_J\beta^{IJKL} \epsilon^{ikl}[E_I^T\delta E_K]_{ik}E^a_{L\, l}=0,
\end{align}
where the lapse-linearity is manifest. 

It will be convenient to decompose $\beta^{IJKL}$ using the symmetric outer product decomposition \cite{comon2008}, i.e. any totally symmetric tensor can be written as,
\begin{align}
\label{sym_outer_prod}
    \beta^{IJKL} = \sum_{r=1}^{R} c_r\,\beta^I_r\,\beta^J_r\,\beta^K_r\,\beta^L_r,
\end{align}
where the $\Ncal$-dimensional vectors $\beta^I_r$ are pairwise non-proportional but not necessarily linearly independent and $R$ is called the symmetric rank. Using this, the Lorentz constraints \eqref{reduced_LC} can be written,
\begin{align}
    \Ccal^a_{I}&= \sum_r \sum_J N_Jc_r\beta^I_r\beta^J_r \epsilon^{ikl}\big[E_I^{\T}\delta \textstyle \sum_K \beta^K_r E_K\big]_{ik} \sum_L \beta^L_r E^a_{L\, l} \notag \\
\label{C=Nbetat}
    &= \displaystyle\sum_r \sum_J N_J\beta^J_r t^a_{rI} =0,
\end{align}
where $t^a_{rI}$ is defined by,
\begin{align}
\label{t_def}
    t^{a}_{rI}= c_r \beta^I_r\epsilon^{ijk} \big[ E^T_I \delta U_r]_{ij}U^a_{r\, k}, && U^a_{r\, i} = \sum_I \beta^I_r E^a_{I\, i}.
\end{align}
A priori, the Lorentz constraints \eqref{reduced_LC} yield $3\Ncal$ equations, but since $\beta^{IJKL}$ is totally symmetric and $\epsilon^{ijk}(E_I^{\T}\delta E^{\,}_J)_{ij}$ is antisymmetric in $IJ$, the sums,
\begin{align}
\label{sum_t=0}
    \sum_I \Ccal^{a}_I =0, && \sum_I t^a_{rI} = c_r \epsilon^{ijk} \big[ U^{\T}_r \delta U^{}_r]_{ij}U^a_{r\, k} =0,
\end{align}
vanish identically, so that $\Ccal^a_I =0$ constitutes only $3(\Ncal-1)$ independent constraints for the $3(\Ncal-1)$ rotational fields $\Omega_I$. The overall rotation is left undetermined by the rotational invariance of the theory.

As discussed in the previous section, for the theory to be ghost-free it is necessary that the Lorentz constraints can be solved without introducing lapse dependence into $\Ccal^I=0$. We formalise this as follows:

\medskip
\noindent\textbf{Lapse-independent solutions:}
We say that $\beta^{IJKL}$ admits lapse-independent solutions if, for generic background vielbeins $\bar{E}_I$, there exist rotations $\Omega^{\mathrm{sol}}_I(\bar{E})$ such that,
\begin{align}
    \mathcal{C}^{a}_I\big(\Omega^{\mathrm{sol}}\bar{E},\,N\big)=0,    
\end{align}
for all $N_I>0$.
\medskip

Since \eqref{C=Nbetat} is linear in $N_I$, lapse-independence is equivalent to the vanishing of each coefficient,
\begin{align}
\label{Xvanish}
    X^a_{IJ}=\sum_r \beta^J_r t^a_{rI} = 0.
\end{align}
However, while $\Ccal^a_I=0$ constitute $3(\Ncal-1)$ constraints, $X^a_{IJ}=0$ yields $3\Ncal(\Ncal-1)$ equations for the same $3(\Ncal-1)$ rotational variables, potentially overconstraining the system. Lapse-independent solutions therefore require an $\Ncal$-fold redundancy in the conditions \eqref{Xvanish}.

In particular, consider the equations for $J=1$,
\begin{align}
    X^a_{I1}=\sum_r \beta^{1}_r t^a_{rI} =0.
\end{align}
These are polynomial in the vielbein components with some constant coefficients given by $\beta^{I1KL}$, and solving these for the rotations $\Omega_I$, they become functions of the spatial vielbeins $\bar{E}_I$ and the constant parameters $\beta^{I1KL}$. 

For $J=2$, we instead get the equations,
\begin{align}
    X^a_{I2}=\sum_r \beta^{2}_r t^a_{rI} =0, 
\end{align}
which again are polynomial in the same variables with the same structure, but with different coefficients $\beta^{I2KL}$. So with the solutions $\Omega^{\mathrm{sol}}_I$, the functions $X^a_{I2}(\Omega^{\mathrm{sol}}(\bar{E}))$ does not identically vanish unless the parameters $\beta^{I1KL}$ are related to $\beta^{I2KL}$, which is not true for generic parameters.\footnote{There certainly are specific vielbein configurations where this does not yield any non-trivial conditions, e.g. if the vielbeins are all proportional $E_I \propto E$, then each term $\epsilon^{ikl}(E_I^{\T}\delta E^{}_K)_{ik}$ is identically zero.} Our goal is to use these relations to restrict $\beta^{IJKL}$ to those which admit the required $\Ncal$-fold redundancy.

\subsection{Irreducible Mutli-gravity interactions}

\begin{figure}
    \centering
\begin{tikzpicture}[scale=1.0]
\tikzstyle{every node}=[draw,shape=circle,fill=black,inner sep=2pt]

\path (0,0) node (c) {};

\node[draw,shape=circle,fill=black!20,minimum size=1.5cm] (Lblob) at (-4,0) {};
\node[draw,shape=circle,fill=black!20,minimum size=1.5cm] (Rblob) at ( 4,0) {};

\draw (c) -- (Lblob);
\draw (c) -- (Rblob);

\def\r{1.85}

\foreach \angle/\name in {
40/l1,
86.667/l2,
133.333/l3,
180/l4,
226.667/l5,
273.333/l6,
320/l7} {
    \path (-4,0) ++(\angle:\r) node (\name) {};
    \draw (Lblob) -- (\name);
}

\foreach \angle/\name in {
140/r1,
93.333/r2,
46.667/r3,
0/r4,
313.333/r5,
266.667/r6,
220/r7} {
    \path (4,0) ++(\angle:\r) node (\name) {};
    \draw (Rblob) -- (\name);
}

\tikzstyle{every node}=[]
\node[] at (0,-0.42) {$e_1$};
\node[] at (-4,0) {$e_I$};
\node[] at (4,0) {$e'_{I'}$};

\end{tikzpicture}
    \caption{Each node $\bullet$ represents a vielbein and edges indicate direct interactions. The vielbein $e_1$ is the only link between the two sectors $\{e_I\}$ and $\{e'_{I'}\}$, shown as grey circles, which otherwise do not interact directly.}
    \label{fig:reducible_sectors}
\end{figure}
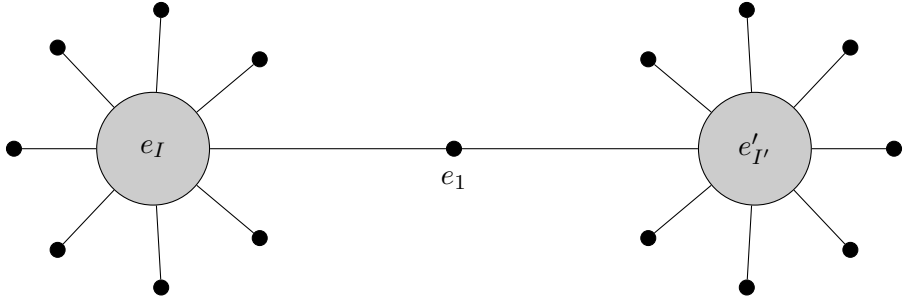

As discussed in Section~\ref{sec:bi_pairwise}, multi-gravity theories can be constructed from pairwise bimetric interactions. We wish to distinguish such extensions from genuine multi-field interactions in \eqref{pot} where all vielbeins couple directly. We also exclude other ``reducible'' interactions, such as sums of determinants \eqref{det_sum_potential}, in which two otherwise independent sectors share at most one vielbein, as illustrated in Figure~\ref{fig:reducible_sectors}. In Section~\ref{sec:generic_trees} we show that such reducible interactions can be built from more fundamental irreducible building blocks.

We now state the assumptions on $\beta^{IJKL}$ under which the following argument applies. We require $\Ncal\geq 3$ and that every pair of vielbeins interacts through at least one term involving a third distinct vielbein. This ensures that the interaction is genuinely multi-field and cannot be decomposed into independent pairwise or single-vielbein-sharing sectors. Explicitly the assumption reads,

\medskip
\noindent\textbf{Irreducible interaction $\Ncal\geq3$} ($\beta^{IJKL}$):
For every pair $I\neq J$ there exist $K,L$ such that $\beta^{IJKL}\neq 0$ with at least one of $K,L$ distinct from $I$ and $J$.
\medskip

\noindent The irreducible interaction assumption can be recast as a condition on the $\beta^I_r$ in the symmetric outer product decomposition \eqref{sym_outer_prod}.

\medskip
\noindent\textbf{Irreducible interaction $\Ncal\geq3$} ($\beta^I_r$):
For every pair $I\neq J$ there exist $r$ such that $\beta^I_r \neq 0$, $\beta^J_r\neq0$ and at least one $K\not\in \{I,J\}$ with $\beta^K_r\neq0$.
\medskip

\noindent The above not only rule out pairwise bimetric interactions, but also the standard bimetric interaction between two fields. Since the bimetric potential cannot meaningfully be reduced to simpler interactions, we also define the bimetric interaction \eqref{bi_pot} to be \textbf{irreducible}, and the arguments in \cite{Comelli:2013txa, Molaee:2018brt, Flinckman:2026xx} provide its uniqueness for $\Ncal=2$. 

With these assumptions in place, we derive restrictions on $\beta^{IJKL}$ for irreducible interactions from the requirement of lapse-independent rotation solutions, which is necessary for the theory to be ghost-free.

\subsubsection{Lemma: Obstructions to \texorpdfstring{$t^a_{rI}=0$}{t=0}}

Before deriving conditions on $\beta^{IJKL}$, we show a result which will become useful later. For irreducible interactions with $R\geq 2$, imposing the conditions $t^a_{rI}=0$ \eqref{t_def} for all $r$ do not provide a viable solution to the Lorentz constraints $\Ccal^a_I=0$ \eqref{reduced_LC} (or $X_{IJ}^a= 0$ \eqref{Xvanish}) for generic vielbeins. 

First, consider a fixed $r$, say $r=1$. The condition $t^a_{1I}=0$ for all $I$ with $\beta^I_1\neq 0$ is equivalent to the symmetrisation conditions,
\begin{align}
\label{sym_cond}
    t^{a}_{1I}= c_1 \beta^I_1\epsilon^{ijk} \big[ E^T_I \delta U_1]_{ij}U^a_{1\, k} \qquad \Longrightarrow \qquad  [E_I^{\T}\delta\, U_1]_{[ij]} = 0,
\end{align}
since $U_1 = \sum_J\beta^J_1 E_J$ is invertible for generic vielbeins. For all $E_I$ with $\beta^I_1\neq 0$ these conditions are non-trivial for generic vielbeins. Each such condition constitutes three equations for the three rotational parameters of $\Omega_I$, so together with the identity $\sum_I t^a_{1I}=0$ \eqref{sum_t=0}, the system \eqref{sym_cond} determines the relative rotations of all vielbeins $E_I$. Since for all these vielbeins, $\beta^I_1\neq 0$, they all appear in $U_1$, so $t^a_{1I}=0$ determines all the relative rotations in the sum $U_1=\sum_J \beta^J_1E_J$, leaving one residual overall rotation.

Now suppose $R\geq 2$ and consider a vielbein $E_I$ with
$\beta^I_1\neq 0$ and $\beta^I_{r}\neq 0$, $r\neq 1$.  Its rotation $\Omega_I$ must simultaneously satisfy,
\begin{align}
\label{first_sys}
    [E_I^{\T}\delta\, U_1]_{[ij]} &= 0,\\
\label{second_sys}
    [E_I^{\T}\delta\, U_{r}]_{[ij]} &= 0\,.
\end{align}
Since $\beta^I_1$ and $\beta^I_{r}$ are non-proportional by construction, the composite vielbeins $U_1$ and $U_{r}$ are generically non-proportional, making these two \emph{independent} equations for the same unknowns.\footnote{Although non-proportional decomposition vectors $\beta_r^I$ generically lead to non-proportional composite vielbeins $U_r$, accidental proportionality can occur on special cases. For instance, if all spatial vielbeins are equal, $E^a_{I\,i}=E^a_i$, then every $U^a_{r \,i} \propto E^a_i$. Such coincidences are non-generic.} The first system \eqref{first_sys} already determines the rotational freedom of $\Omega_I$ and the second \eqref{second_sys} imposes three additional conditions that generically cannot be satisfied without restricting the non-rotational fields $\bar{E}_I$. We will now argue that for irreducible interactions with $R\geq 2$, at least one such ``doubly constrained'' vielbein must exist if we impose $t^a_{rI}=0$.  

First, call a vielbein $E_{I'}$ a \emph{leaf} if $\beta^{I'}_r\neq 0$ for exactly one value of $r=r_\star$. A leaf vielbein must then be symmetrised with only one $U_{r_\star}$, $[E^{\T}_{I'}\delta U_{r_\star}]_{[ij]}=0$ and will not be overconstrained. We claim that if every vielbein were a leaf, irreducible interactions would necessarily have $R=1$, contradicting the assumption $R\geq 2$. 
 
To see this, suppose every vielbein is a leaf. Without loss of generality, let $E_1$ be a leaf with its unique nonzero factor at $r=1$, so that $\beta^1_1\neq 0$. The irreducible interaction assumption requires that every other vielbein $E_J$ interacts with $E_1$, i.e. there exists some $r$ with $\beta^{1}_r\neq 0$ and $\beta^J_r\neq 0$ for all $J$. Since $E_1$ is a leaf, the only non-zero $\beta^{1}_r$ is $\beta^1_1$ ($r=1$), which forces $\beta^J_{1}\neq 0$ for every $J$. Since $E_J$ is also a leaf, its unique nonzero factor must be $\beta^J_1$ as well, so we find that $\beta^J_{r\neq1} = 0$ for every $J$. The decomposition then has only one nonzero term in \eqref{sym_outer_prod}, giving $R=1$, contradicting the $R\geq2$ assumption. Therefore, for $R\geq2$ there will always exist at least one vielbein which is not a leaf.

Since every vielbein has $\beta^I_1\neq 0$ in an irreducible interaction where $E_1$ is a leaf, all vielbeins $E_I$ appear in $U_1= \sum_I \beta^I_1 E_I$. $\beta^I_1\neq 0$ also implies that $t^a_{1I}=0$ is non-trivial, so every vielbein must be symmetrised with $U_1$, and the system,
\begin{align}
\label{leaf_system}
    t^a_{1I}=c_1\beta^I_1 \epsilon^{ijk}[E_I^{\T}\delta U_1]_{ij}U^a_{1\, k}=0 \qquad \Longrightarrow \qquad [E_I^{\T}\delta U_1]_{[ij]}=0, \qquad \forall \, I
\end{align}
is enough to determine all the relative rotations $\Omega_I$.

Since $R\geq 2$, at least one vielbein is not a leaf. Without loss of generality, let this be $E_2$, with $\beta^2_{r}\neq 0$ for some $r\neq 1$. Since $\beta^2_1\neq 0$, the rotation $\Omega_2$ is already fully determined by the condition $[E_2^{\T}\delta U_1]_{[ij]}=0$, yet $t^a_{r\,2}=0$ implies it must additionally satisfy $[E_2^{\T}\delta U_{r}]_{[ij]}=0$. Since $U_{r}$ is generically non-proportional to $U_1$, this is an independent condition on an already fixed $\Omega_2$.

One might hope that the identity $\sum_J t^a_{rJ}=0$ (effectively corresponding to the overall rotation) could render the additional condition redundant. Indeed, this is precisely the mechanism that makes the Lorentz constraints in pairwise bimetric and other reducible interactions lapse-independent, as we will see explicitly in Section~\ref{sec:generic_trees}. The idea is as follows: if all other vielbeins $E_{J\neq 2}$ appearing in $U_{r}$ can independently be solved so that $t^a_{rJ}=0$, the identity,
\begin{align}
\label{leaf_identity}
    \sum_J t^a_{r J}= t^a_{r\,2} + \mathrm{(vanishing\; terms)}=0
\end{align}
would force $t^a_{r\,2}=0$ automatically, without imposing a new condition on $\Omega_2$. However, this mechanism requires that the vielbeins $E_{J\neq 2}$ in $U_r$ have been symmetrised with $U_r$. But the leaf $E_1$ necessarily interacts with $E_J$ via $U_1$, so the rotations of each $E_J$ have already been fully determined by the system \eqref{leaf_system}, which enforces $[E_J^{\T}\delta U_1]_{[ij]}=0$, not $[E_J^{\T}\delta U_r]_{[ij]}=0$, therefore $t^a_{rJ}\neq0$. For the identity \eqref{leaf_identity} to nevertheless render $t^a_{r\,2}=0$ redundant, the symmetrisation with $U_1$ would have to imply symmetrisation with $U_r$ as well. This requires $U_r \propto U_1$, which for generic vielbeins holds only if $\beta^I_r\propto \beta^I_1$, contradicting the non-proportionality of the decomposition vectors. The extra condition $[E^{\T}_2\delta U_r]_{[ij]}=0$ is therefore independent, and $\Omega_2$ is overconstrained.

If there are no leaves, then $t^a_{rI}=0$ is non-trivial for at least two $r$ for every $I$ and thus implies that every vielbein must be symmetrised with more than one composite vielbein $U_r$, again overconstraining it.

So $t^a_{rI}=0$ for all $r$ is not a viable solution for generic vielbeins whenever $R\geq 2$ and the interaction is irreducible. Any $\Ncal$-fold redundancy in the conditions $X^a_{IJ}=0$ \eqref{Xvanish} must therefore arise from the structure of $\beta^{IJKL}$ itself, rather than from the individual vanishing of each $t^a_{rI}$. We now turn to deriving what this implies for the form of $\beta^{IJKL}$.

\subsubsection{Warm-up: \texorpdfstring{$R=2$}{R=2}}
\label{sec:warmup}
We start by showing that an irreducible multi-gravity interaction cannot have symmetric rank $2$. For $R=2$, the conditions $X^a_{IJ}=0$ read,
\begin{align}
\label{X_R=2}
    X^a_{IJ} = \beta^J_1t^a_{1I} + \beta^J_2t^a_{2I} = 0,
\end{align}
and for some fixed $J$, say $J=1$, we can solve $X^a_{I1}=0$ for the rotations $\Omega^{\mathrm{sol}}_I(\bar{E}, \beta^1)$. For $J'\neq 1$,
\begin{align}
    X^a_{IJ'}= \beta^{J'}_1t^a_{1I}(\Omega^{\mathrm{sol}}) + \beta^{J'}_2t^a_{2I}(\Omega^{\mathrm{sol}}),
\end{align}
is generically non-zero unless the coefficients are proportional, forcing $t^a_{1I}=t^a_{2I}=0$, which we have shown to be inconsistent.

To make this precise, for fixed $I$ and $a$ we define the matrix $\mathbb{B}$ and vector $T^a_I$,
\begin{align}
    \mathbb{B} = \begin{pmatrix} \beta^1_1 & \beta^1_2 \\[2pt]
    \beta^2_1 & \beta^2_2 \\
    \vdots & \vdots \\
    \beta^{\Ncal}_1 & \beta^{\Ncal}_2
    \end{pmatrix}, &&T_I^a = \begin{pmatrix} t^a_{1I} \\ t^a_{2I} \end{pmatrix}.
\end{align}
The conditions \eqref{X_R=2} for all $J$ are then equivalent to,
\begin{align}
    \mathbb{B}\,T_I^a = 0.
\end{align}
Since the columns of the $\Ncal\times 2$ matrix $\mathbb{B}$ are non-proportional by assumption, $\mathbb{B}$ has rank $2$ and hence trivial right kernel, so,
\begin{align}
    \mathbb{B}\,T^a_I =0 \qquad \Longrightarrow \qquad T^a_I=0 \qquad \Longrightarrow \qquad t^a_{rI}=0 \qquad \forall \; r, I, a.
\end{align}
But as shown in the previous section, $t^a_{rI}=0$ for all $r$ overconstrains the system for irreducible interactions. Therefore, $\beta^{IJKL}$ with symmetric rank $R=2$ does not admit lapse-independent solutions for $\Ncal\geq3$.

\subsubsection{Linearly independent factors}
The above argument generalises to arbitrary $R$ and $\Ncal$. However, for $R>2$ the vectors $\beta^I_r$ need not be linearly independent, only pairwise non-proportional, which introduces additional redundancy that could in principle allow lapse-independent solutions. Before showing that this never happens, we present the argument for the special case where the $\beta^I_r$ form a linearly independent set, which in particular requires $R\leq\Ncal$.
The conditions $X^a_{IJ}=0$ for all $J$ read,
\begin{align}
    X^a_{IJ}=\sum_{r=1}^R \beta^J_r\,t^a_{rI} = 0.
\end{align}
For fixed $I$ and $a$, define the matrix $\mathbb{B}$ and vector $T^a_I$,
\begin{align}
    \mathbb{B} = 
    \begin{pmatrix}
        \beta^1_1 & \cdots & \beta^1_R \\
        \vdots & & \vdots \\
        \beta^{\Ncal}_1 & \cdots & \beta^{\Ncal}_R
    \end{pmatrix}, && T^a_I = \begin{pmatrix} t^a_{1I} \\ \vdots \\ t^a_{RI} \end{pmatrix}.
\end{align}
Then $X^a_{IJ}=0$ for all $J$ is equivalent to,
\begin{align}
    \mathbb{B}\,T^a_I = 0.
\end{align}
Since the columns of the $\Ncal\times R$ matrix $\mathbb{B}$ are linearly independent by assumption, $\mathbb{B}$ has rank $R$ and hence trivial right kernel, so,
\begin{align}
    \mathbb{B}\,T^a_I =0 \qquad \Longrightarrow \qquad T^a_I=0 \qquad \Longrightarrow \qquad t^a_{rI}=0 \qquad \forall \; r, I, a,
\end{align}
which as shown above overconstrains the system for irreducible interactions with $R\geq 2$. We conclude that linearly independent decomposition vectors $\beta^I_r$ are incompatible with lapse-independent solutions for irreducible interactions with $R\geq2$.

\subsubsection{Linearly dependent}

We now extend the argument to arbitrary symmetric rank $R\geq 2$ without assuming that the decomposition vectors $\beta^I_1,\dots,\beta^I_R$ are linearly independent. The additional redundancy can be handled by considering the linear span $S$ of the $R$ vectors $\beta^I_r$, i.e.\ $S$ is the space of all linear combinations of $\beta^I_r$. If the span $S$ has dimension $s$ ($1\leq s\leq \min(R,\Ncal)$), we can choose a basis $b^I_1,\dots,b^I_s$ and express $\beta^I_r$ as,
\begin{align}
    \beta^I_r = \sum_{n=1}^s b^I_nA^n_{~r},
\end{align}
for an $s\times R$ matrix $A^n_{~r}$. The rows of $A^n_{~r}$ (indexed by $n$) are by construction linearly independent.\footnote{If a row were a linear combination of the others, say $A^{n'}_{~r}=\sum_{n\neq n'}\mu_n A^n_{~r}$, then $\beta^I_r = \sum_n b^I_n A^n_{~r} = \sum_{n\neq n'}(b^I_n + \mu_n b^I_{n'})A^n_{~r}$, so every $\beta^I_r$ would lie in the span of the $s-1$ vectors $b^I_n + \mu_n b^I_{n'}$, contradicting $\dim S = s$.} Note that since $R\geq2$, we can always choose at least two of the decomposition vectors as basis vectors, say $b^I_1=\beta^I_1$ and $b^I_2=\beta^I_2$, so that $A^1_{~r}=\delta^1_r$ and $A^2_{~r}=\delta^2_r$.

The conditions $X^a_{IJ}=0$ can now be expressed in terms of $b^J_n$,
\begin{align}
    X^a_{IJ}=\sum_{r=1}^R \beta^J_r\,t^a_{rI} =  \sum_{n=1}^s b^J_n\,\theta^a_{n\,I}=0,
\end{align}
where we defined the effective combinations,
\begin{align}
    \theta^a_{n\, I}=\sum_{r=1}^R A^n_{\; r}\,t^a_{rI}.
\end{align}
For fixed $I$ and $a$, define the matrix $\mathbb{B}$ and vector $\Theta^a_I$,
\begin{align}
    \mathbb{B} =
    \begin{pmatrix}
        b^1_1 & \cdots & b^1_s \\
        \vdots & & \vdots \\
        b^{\Ncal}_1 & \cdots & b^{\Ncal}_s
    \end{pmatrix}, && \Theta^a_I = \begin{pmatrix} \theta^a_{1I} \\ \vdots \\ \theta^a_{sI} \end{pmatrix}.
\end{align}
Then $X^a_{IJ}=0$ for all $J$ is equivalent to,
\begin{align}
    \mathbb{B}\,\Theta^a_I = 0.
\end{align}
Since the columns $b^I_n$ of the $\Ncal \times s$ matrix $\mathbb{B}$ are linearly independent by construction, $\mathbb{B}$ has rank $s$ and trivial right kernel. Therefore,
\begin{align}
    \mathbb{B}\Theta^a_I =0 \qquad \Longrightarrow \qquad \Theta^a_I=0 \qquad \Longrightarrow \qquad \theta^a_{nI}=0 \qquad \forall \; n, I, a.
\end{align}
Lapse-independence thus forces $s$ systems of conditions $\theta^a_{nI}=0$ for each $a$ and $I$, yielding $3s(\Ncal-1)$ equations for $3(\Ncal-1)$ unknown rotations. For $s\geq 2$ the system is overconstrained unless the $s$ systems are mutually redundant. We now show that this redundancy cannot occur for irreducible interactions, by treating the two mutually exclusive and exhaustive cases:
\begin{itemize}[leftmargin=2.54cm]
    \item[\textbf{With leaves:}] Suppose there exists a leaf vielbein, say $E_1$ with $\beta^1_1\neq 0$ as its only nonzero factor. Under the irreducible interaction assumption, $\beta^I_1\neq 0$ for all $I$, as shown above. We also showed that not every vielbein can be a leaf, so let $E_2$ be a non-leaf with $\beta^2_2\neq 0$. We choose our basis so that $b^I_1=\beta^I_1$ and $b^I_2=\beta^I_2$, giving $A^1_{~r}=\delta^1_r$ and $A^2_{~r}=\delta^2_r$. The first two $\theta$-systems then reduce to,
    \begin{align}
        \theta^a_{1I} = t^a_{1I} = 0, && \theta^a_{2I} = t^a_{2I} = 0.
    \end{align}
    The first system determines all the rotations $\Omega^{\mathrm{sol}}_I$ as established in \eqref{leaf_system}. For the second system, any leaf $E_{I'}$ with $\beta^{I'}_{r\neq 1}=0$ satisfies $t^a_{2I'}=0$ trivially. However, the non-leaf $E_2$ has $\beta^2_2\neq 0$, so $\theta^a_{22}=t^a_{22}=0$ imposes the symmetrisation condition $[E_2^{\T}\delta U_2]_{[ij]}=0$. Since $\Omega_2$ is already fully determined by $[E_2^{\T}\delta U_1]_{[ij]}=0$ and $U_2$ is generically non-proportional to $U_1$, this overconstrains $\Omega_2$.
    \item[\textbf{No leaves:}] If no leaves exist, every vielbein has $\beta^I_r\neq 0$ for at least two values of $r$. This is precisely the situation shown to be problematic in the obstruction argument: each vielbein must be symmetrised with at least two composite vielbeins $U_r$, and the first symmetrisation already determines its rotation. Moreover, the identity $\sum_I t^a_{rI}=0$ cannot render the second condition redundant, since this mechanism relies on other vielbeins in the sum having $t^a_{rI}=0$ trivially, which required those vielbeins to be leaves. Without leaves, every vielbein contributes non-trivially to both systems, and the overconstraining cannot be avoided.
\end{itemize}
In both cases, lapse-independent solutions cannot exist for $s\geq 2$, and the only remaining possibility is $s=1$, i.e. all decomposition vectors are proportional,
\begin{align}
    \beta_r^I=\lambda_r \beta^I
    \qquad\Longrightarrow\qquad
    \beta^{IJKL}
    =
    \Big(\sum_r c_r \lambda_r^4\Big)\beta^I\beta^J\beta^K\beta^L,
\end{align}
so $\beta^{IJKL}$ is of rank 1 and can be written, up to a constant rescaling, as,
\begin{align}
\label{rank_one_beta}
    \beta^{IJKL}= \beta^I \beta^J \beta^K \beta^L.  
\end{align}
We therefore conclude that the only subset of \eqref{pot} with irreducible multi-gravity interaction and lapse-independent Lorentz constraints is of the form,
\begin{align}
    V &= \frac{2m^4}{4!}\!\!\!\!\sum_{IJKL=1}^\Ncal \!\!\!\!\beta^I\beta^J \beta^K \beta^L\epsilon_{ABCD}\epsilon^{\alpha \beta \gamma \delta}e^A_{I\, \alpha}e^B_{J\, \beta}e^C_{K\, \gamma}e^D_{L\, \delta} = 2m^4 \det(\sum_I \beta^I e_I),
\end{align}
which is precisely the multivielbein theory presented in \cite{Hassan:2018mcw} and further analysed in \cite{Flinckman:2024zpb, Flinckman:2025bje}.

Note that the condition \eqref{rank_one_beta} holds for all components $\beta^{IJKL}$ except $\beta^{IIII}$ which does not enter the Lorentz constraints and is therefore arbitrary. $\beta^{IIII}$ contributes only a cosmological constant for $g_I$, so any part which is not equal to $(\beta^I)^4$ can be parametrised by $\Lambda_I$.

\section{Reducible interactions}
\label{sec:generic_trees}
 
In the previous section we showed that the only irreducible multi-gravity interaction within the Hinterbichler–Rosen class \eqref{pot} that admits lapse-independent Lorentz constraints, and thus potentially can be ghost-free, has the form $\beta^{IJKL}=\beta^I\beta^J\beta^K\beta^L$, which yields precisely the determinant coupling \eqref{det_coupling}. We now turn to more general, reducible interactions built from the two ghost-free irreducible building blocks: the bimetric potential $V_{\mathrm{bi}}$ \eqref{bi_pot} and the determinant coupling $V_{\det}$ \eqref{det_coupling}. These reducible interactions have $\beta^{IJKL}$ of symmetric rank $R\geq 2$, yet admit lapse-independent solutions for the rotations provided the interaction graph forms a tree. Explicit examples, including counterexamples where the tree condition is violated, are presented in Appendix~\ref{app:examples}.

\subsection{Interaction graphs}
\begin{figure}
    \centering
\begin{tikzpicture}[scale=1.0]
\tikzset{
    vnode/.style={draw, circle, fill=black, inner sep=2pt, minimum size=6pt},
    inode/.style={draw, circle, fill=white, inner sep=2.5pt, minimum size=8pt},
}
    \node[vnode, label={below:$e_1$}] (a1) at (-6.5, 0) {};
    \node[inode] (b1) at (-5.5, 0) {};
    \node[vnode, label={below:$e_2$}] (a2) at (-4.5, 0) {};
    \node[inode] (b2) at (-3.5, 0) {};
    \node[vnode, label={below:$e_3$}] (a3) at (-2.5, 0) {};
    
    \draw (a1) -- (b1);
    \draw (b1) -- (a2);
    \draw (a2) -- (b2);
    \draw (b2) -- (a3);
    
    \node at (-1.5, 0) {\Large $=$};
    
    \node[vnode, label={below:$e_1$}] (c1) at (-0.5, 0) {};
    \node[vnode, label={below:$e_2$}] (c2) at (0.5, 0) {};
    \node[vnode, label={below:$e_3$}] (c3) at (1.5, 0) {};
    
    \draw (c1) -- (c2);
    \draw (c2) -- (c3);
    
    \node at (-2.5, -0.9) {\textbf{a)}};
    
    \node[inode, label={above:$\Acal$}] (detA) at (5, 0.7) {};
    
    \node[vnode, label={below:$e_1$}] (e1) at (3.8, -0.5) {};
    \node[vnode, label={below:$e_2$}] (e2) at (5, -0.5) {};
    \node[vnode, label={below:$e_3$}] (e3) at (6.2, -0.5) {};
    \node[vnode, label={below:$e_4$}] (e4) at (7.5, -0.5) {};
    
    \draw (e1) -- (detA);
    \draw (e2) -- (detA);
    \draw (e3) -- (detA);
    \draw (e3) -- (e4);
    
    \node at (5.65, -1.2) {\textbf{b)}};
\end{tikzpicture}
    \caption{Filled vertices ($\bullet$) represent vielbeins, open vertices ($\circ$) represent determinant sectors. \textbf{a)} For bimetric interactions, the $\circ$ is suppressed and a direct edge $\bullet$--$\bullet$ is drawn instead. \textbf{b)} A determinant-plus-bimetric interaction with $\Acal=\{1,2,3\}$ and a bimetric edge connecting $e_3$ and $e_4$.}
    \label{fig:convention_and_det_bi}
\end{figure}
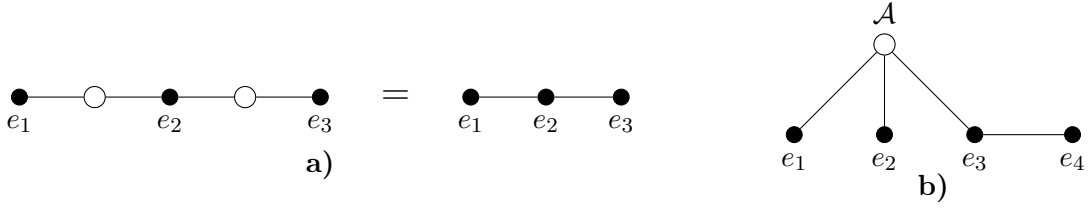

To describe the structure of reducible interactions precisely, we extend the notion of interaction graphs used in previous literature, e.g. \cite{Hinterbichler:2012cn,Nomura:2012xr,Afshar:2014dta, Scargill:2014wya} to include the determinant interactions. We represent the interaction as a \emph{bipartite} graph $\Gcal$ with two types of vertices:
\begin{itemize}
    \item[--] \textbf{Vielbein vertices} $\bullet$: one for each vielbein $e_I$.
    \item[--] \textbf{Interaction vertices} $\circ$: one for each determinant or bimetric potential.
\end{itemize}
An edge $-$ connects a vielbein vertex $\bullet$ to an interaction vertex $\circ$ whenever $e_I$ participates in that interaction. Since a bimetric interaction connects exactly two vielbeins, its $\circ$ can be suppressed, writing $\bullet$--$\bullet$ as shorthand for $\bullet$--$\circ$--$\bullet$, consistent with the pairwise notation used in the literature, see Figure~\ref{fig:convention_and_det_bi}a. With this convention, visible $\circ$ vertices represent determinant interactions. A vielbein appearing in only one interaction is a \emph{leaf} of $\Gcal$, connected to a single $\circ$ or $\bullet$, in agreement with our previous convention of a leaf. In Figure \ref{fig:convention_and_det_bi}b, $e_1,e_2$ and $e_4$ are leafs, while $e_3$ is not.
 
\subsection{Generic interaction trees}

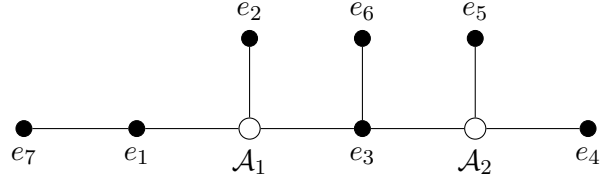
\begin{figure}
    \centering
\begin{tikzpicture}[scale=1.0]
\tikzset{
    vnode/.style={draw, circle, fill=black, inner sep=2pt, minimum size=6pt},
    inode/.style={draw, circle, fill=white, inner sep=2.5pt, minimum size=8pt},
}
    \node[vnode, label={below:$e_7$}] (e7) at (-4.5, 0) {};
    \node[vnode, label={below:$e_1$}] (e1) at (-3, 0) {};
    \node[inode, label={below:$\Acal_1$}] (A1) at (-1.5, 0) {};
    \node[vnode, label={below:$e_3$}] (e3) at (0, 0) {};
    \node[inode, label={below:$\Acal_2$}] (A2) at (1.5, 0) {};
    \node[vnode, label={below:$e_4$}] (e4) at (3, 0) {};

    \node[vnode, label={above:$e_2$}] (e2) at (-1.5, 1.2) {};
    \node[vnode, label={above:$e_5$}] (e5) at (1.5, 1.2) {};
    \node[vnode, label={above:$e_6$}] (e6) at (0, 1.2) {};

    \draw (e7) -- (e1);
    \draw (e1) -- (A1);
    \draw (e2) -- (A1);
    \draw (e3) -- (A1);
    \draw (e3) -- (e6);
    \draw (e3) -- (A2);
    \draw (e5) -- (A2);
    \draw (e4) -- (A2);
\end{tikzpicture}
    \caption{A mixed interaction tree on seven vielbeins. The determinant sectors $\Acal_1=\{e_1,e_2,e_3\}$ and $\Acal_2=\{e_3,e_4,e_5\}$ share the vielbein $e_3$, which also interacts with $e_6$ via a bimetric edge. A second bimetric edge connects $e_1$ to $e_7$.}
    \label{fig:mixed_tree}
\end{figure}

The most general potential built from the irreducible building blocks takes the form,
\begin{align}
\label{general_tree_potential}
    V = \sum_{r} V_{\mathrm{det}}(u_r) + \sum_{(I'J') \in \mathcal{T}} V_{\mathrm{bi}}(e_{I'},e_{J'})\,, && u_r = \sum_{I\in \Acal_r}\beta^I_r e_I,
\end{align}
where $\mathcal{T}$ denotes the set of bimetric edges and $\Acal_r$ the index set of vielbeins in the determinant interaction $V_{\mathrm{det}}(u_r)$. The full $\beta$-parameter for such an interaction takes the form,
\begin{align}
\label{gen_int_beta}
    \beta^{IJKL} = \sum_{r} \beta^I_r\beta^J_r\beta^K_r\beta^L_r + \sum_{(I'J') \in \mathcal{T}} \beta^{IJKL}_{(I'J')}\,,
\end{align}
where $\beta^I_r$ is nonzero only for $I\in \Acal_r$ and $\beta^{IJKL}_{(I'J')}$ is nonzero only when all indices are $I'$ or $J'$, but otherwise arbitrary, admitting the representation \eqref{bi_beta}.

The interaction \eqref{general_tree_potential} can be represented by an interaction graph $\Gcal$ as described above. The structural requirement for lapse-independent Lorentz constraints is that $\Gcal$ is a \emph{tree}, i.e.\ connected and cycle-free, see for example Figure~\ref{fig:mixed_tree}. This implies that any two determinant interactions share at most one vielbein and that there are no bimetric cycles. It also guarantees that there exist at least two leaves and precisely $\Ncal-1$ edges.
\newpage

The Lorentz constraints can then be solved by the following leaf-removal procedure on $\Gcal$:
\begin{enumerate}
    \item Identify all leaf vielbeins of $\Gcal$. For each leaf $e_{I'}$:
    \begin{enumerate}
        \item[a)] If $e_{I'}$ is attached to a determinant sector $\Acal_r$: its Lorentz constraints reduces to exactly one symmetrisation condition,
        \begin{align}
        \label{det_leaf}
            [e_{I'}^{\T}\eta\, u^{\,}_r]_{[\mu \nu]}=0,    
        \end{align}
        which determines $\Omega_{I'}$. Remove $e_{I'}$ from $\Gcal$ and from $\Acal_r$.
        \item[b)] If $e_{I'}$ is attached to another vielbein $e_{J'}$ via a bimetric edge: its Lorentz constraints reduces to,
        \begin{align}
        \label{bim_leaf}
            [e_{I'}^{\T}\eta\, e^{\,}_{J'}]_{[ij]}=0,    
        \end{align}
        which determines $\Omega_{I'}$. Remove $e_{I'}$ and the edge from $\Gcal$.
    \end{enumerate}
    \item For any determinant sector $\Acal_r$ now left with a single vielbein, the identity, 
    $$\sum_{I\in\Acal_r}\beta^I[e_I^{\T}\eta u_r]_{[\mu \nu]}=0,$$ 
    ensures that $\Acal_r$ contributes nothing to the remaining vielbein's Lorentz constraints. Remove the vertex $\Acal_r$ from $\Gcal$. The result is a smaller tree.
    \item Repeat from step 1 until a single vielbein remains. Its equation is automatically satisfied by the global identity $\sum_I \Ccal^a_{I}=0$.
\end{enumerate}
At each step, every condition reduce to symmetrisation conditions of the form $\eqref{det_leaf}$ or \eqref{bim_leaf}, so that the corresponding vielbein is only constrained by one equation. As the tree shrinks, any vielbein which previously was not a leaf, but now is, is not constrained by the previous leaf structure as those equations have been solved by the leaves and the resulting contribution is now zero. Since a tree always have $\Ncal-1$ edges, this guarantee that all the $6(\Ncal-1)$ relative Lorentz fields are determined.

Conversely, if $\Gcal$ contains a cycle, the leaf-removal procedure terminates before all vielbeins have been eliminated. The remaining vielbeins each participate in at least two interactions, and the arguments of Appendix~\ref{beta_1234} and Appendix~\ref{sec:pairwise_bimetric} show that the conditions $X^a_{IJ}=0$ overconstrain the rotations.

For $\Ncal=2$, the only interaction graph is trivially a tree, but for $\Ncal=3$, the only reducible interaction graphs are all pairwise bimetric interactions, either the line (see Figure \ref{fig:bimetric_trees}a) or 3-cycle (see Figure \ref{fig:bimetric_cycle}). Of these, only the line ($\bullet$–$\bullet$–$\bullet$) admits lapse-independent solutions and corresponds to a tree, while the $3$-cycle does not. We therefore conjecture that the most general multi-gravity theory within the Hinterbichler–Rosen class \eqref{pot} that admits lapse-independent Lorentz constraints are combinations of bimetric and determinant interactions, where the interaction graph $\Gcal$ is a tree.

\subsection{Trees of bimetric potential}
If the tree only contains bimetric interactions, the leaf-removing algorithm above will guarantee that all the Lorentz constraints reduce to pairwise symmetrisation \eqref{pairwise_sym},
\begin{align}
    e^{\T}_I \eta e_J^{\;} =e^{\T}_J \eta e_I^{\;} 
\end{align}
for all $(IJ)\in \Gcal$. Since $\Gcal$ is a tree, this is $6(\Ncal-1)$ conditions on the equally many Lorentz fields. As argued in the end of Section \ref{sec:rev_bimetric}, \eqref{pairwise_sym} guarantees the existence of a real square root,
\begin{align}
    (e_I^{-1}e_J) = \sqrt{g_I^{-1}g_J},
\end{align}
and the theory has a metric equivalent formulation,
\begin{align}
    V = 2m^4 \sum_{(IJ)\in \Gcal} \sqrt{g_I}\sum_{n=0}^4 P_n(\sqrt{g_I^{-1}g_J}).
\end{align}
In Appendix \ref{sec:pairwise_bimetric} we show this procedure explicitly in the shift-less Ansatz, and also how this fails if the pairwise interactions form a cycle, where the metric and vielbein formulations are no longer equivalent.

\section{Conclusion and discussions}
\label{sec:conclusion}

Our main conclusion is that, within the general class of antisymmetrised multivielbein interactions introduced in \cite{Hinterbichler:2012cn}, the determinant interaction found in \cite{Hassan:2018mcw} is the unique ghost-free theory of multiple spin-2 fields exhibiting genuine irreducible multi-field interactions. Technically, we show that necessary conditions on the absence of ghosts restrict the general symmetric multivielbein couplings $\beta^{IJKL}$ of \cite{Hinterbichler:2012cn} to have {\it symmetric tensor rank} $R=1$, that is, $\beta^{IJKL}=\beta^I \beta^J \beta^K \beta^L$. Our results are summarized in more detail in Section \ref{summary}. 

While our main result is about the uniqueness of irreducible multivielbein interactions, as specified in Section~\ref{sec:restric}, we have also described the construction of reducible ghost-free interactions, using irreducible multimetric and bimetric blocks, arranged in interaction trees. We have not proven that these interaction trees exhaust all ghost-free reducible interactions. In principle, there may exist reducible interactions satisfying the necessary ghost-free conditions that are not captured by our interaction trees, but we have not managed to find any such examples. The reducible interactions that we can find not covered by our algorithm are not ghost-free (for example, a closed loop of bimetric interactions). Hence we conjecture that the interaction trees described here cover all possible reducible ghost-free theories.  

While in this work we have investigated ghost-free multivielbein theories within the framework of \cite{Hinterbichler:2012cn}, one may ask if ghost-free theories could exist outside of this framework. We will address this question in an upcoming paper \cite{Flinckman:2026xx}.

\newpage
\appendix
\section{Examples of reducible interactions}
\label{app:examples}

In this appendix we present explicit examples of how the irreducible building blocks, bimetric and determinant interactions, can be combined into reducible interactions. We verify in each case whether the conditions $X^a_{IJ}=0$ admit lapse-independent solutions, illustrating the leaf-removal procedure of Section~\ref{sec:generic_trees} and its failure in the presence of cycles. While the analysis can be done for fully covariant vielbeins, we will use shiftless ansatz \eqref{mini_ansatz} to align with the notation in the rest of the paper. We also define,
\begin{align}
\label{M_def}
    M^i_{IJ} = \tfrac{1}{2}\epsilon^{ijk}(E_I^{\T} \delta\, E_J)_{jk},
\end{align}
where $M^i_{IJ}=-M^i_{JI}$, $M^i_{II}=0$, so that,
\begin{align}
    X^a_{IJ}(E) = 2\sum_{KL}\beta^{IJKL}\,M^i_{IK}(E)\,E^a_{L\,i}.
\end{align}
The condition $M^i_{IJ}=0$ is equivalent to the symmetrisation condition $[E_I^{\T}\delta E_J]_{[ij]}=0$.

\subsection{Counterexample: the \texorpdfstring{$\beta^{1234}$}{ß1234}-interaction}
\label{beta_1234}

We begin with a simple irreducible interaction that fails to admit lapse-independent solutions.\footnote{A similar interaction was considered in 2+1 dimensions in \cite{Afshar:2014dta}.} Consider $\beta^{IJKL}$ with only $\beta^{1234}\neq 0$, which we can set to $1/2$. Despite its simplicity, $\beta^{IJKL}$ has symmetric rank 8.\footnote{Up to normalisation, one decomposition is $\beta^I_1=(1,1,1,1)$, $\beta^I_2=(1,1,1,-1)$, $\beta^I_3=(1,1,-1,1)$, $\beta^I_4=(1,1,-1,-1)$, $\beta^I_5=(1,-1,1,1)$, $\beta^I_6=(1,-1,1,-1)$, $\beta^I_7=(1,-1,-1,1)$, $\beta^I_8=(1,-1,-1,-1)$.} The potential reduces to,
\begin{align}
    V= m^4\,\epsilon_{ABCD}\epsilon^{\alpha\beta\gamma\delta}e^A_{1\,\alpha}e^B_{2\,\beta}e^C_{3\,\gamma}e^D_{4\,\delta}.
\end{align}
Note that in the interaction graph language there are no leaves, and so the leaf removing algorithm cannot be employed.

For $I=1$, the conditions $X^a_{1J}=0$ read,
\begin{subequations}
\label{4-wedge_X}
\begin{align}
    X^a_{12} &= M^i_{13}E^a_{4\,i}
    + M^i_{14}E^a_{3\,i}\, = 0, \\
    X^a_{13} &= M^i_{12}E^a_{4\,i}
    + M^i_{14}E^a_{2\,i} \,= 0, \\
    X^a_{14} &= M^i_{12}E^a_{3\,i}
    + M^i_{13}\,E^a_{2\,i} = 0.
\end{align}    
\end{subequations}
These are 9 equations for the 3 rotational parameters $\Omega_1$, involving $M^i_{12}$, $M^i_{13}$ and $M^i_{14}$. The equations for $I=2,3,4$ follow by permutation of the indices.

Since the spatial vielbeins are invertible, we can multiply the first two equations by $E^j_{4\, a}$ to solve for $M^i_{12}$ and $M^i_{13}$, and substitute into the third to obtain,
\begin{align}
    X^a_{14}=- \big[E_3 E_4^{-1}E_2 + E_2E^{-1}_4E_3 \big]^a_{~i}M^i_{14}=0.
\end{align}
Since $E_3 E_4^{-1}E_2 + E_2E^{-1}_4E_3$ is invertible for generic vielbeins, the only solution is $M^i_{14}=0$, which also implies the conditions $M^i_{12}=M^i_{13}=0$, 
\begin{align}
    [E_1^{\T}\delta E_2]_{[ij]}=0, \qquad [E_2^{\T}\delta E_3]_{[ij]}=0,\qquad [E_3^{\T}\delta E_1]_{[ij]}=0,
\end{align}
which are $3\times3=9$ independent conditions on $2\times 3=6$ relative rotations. The system is therefore overconstrained for generic vielbeins, confirming that the $\beta^{1234}$-interaction does not admit lapse-independent solutions.

\subsection{Pairwise bimetric interactions}
\label{sec:pairwise_bimetric}

\subsubsection{\texorpdfstring{$\Ncal=3$}{N=3} pairwise bimetric interaction}

\begin{figure}
    \centering
\begin{tikzpicture}[scale=1.0]
\tikzstyle{every node}=[draw,shape=circle,fill=black,inner sep=2pt]

\path (-4,0) node (a1) {};
\path (-2,0) node (a2) {};
\path ( 0,0) node (a3) {};
\draw (a1) -- (a2);
\draw (a2) -- (a3);
\tikzstyle{every node}=[]
\node at (-4,-0.4) {$e_1$};
\node at (-2,-0.4) {$e_2$};
\node at ( 0,-0.4) {$e_3$};
\node at (-2,-1.2) {\textbf{a)}};

\tikzstyle{every node}=[draw,shape=circle,fill=black,inner sep=2pt]
\path (4,0) node (b1) {};
\path (6,0) node (b2) {};
\path (8,0) node (b3) {};
\path (6,1.5) node (b4) {};
\path (8,1.5) node (b5) {};
\path (9.5,0) node (b6) {};
\path (8,-1.5) node (b7) {};
\draw (b1) -- (b2);
\draw (b2) -- (b3);
\draw (b2) -- (b4);
\draw (b3) -- (b5);
\draw (b3) -- (b6);
\draw (b3) -- (b7);
\tikzstyle{every node}=[]
\node at (4,-0.4) {$e_1$};
\node at (6.4,0.4) {$e_2$};
\node at (8.4,0.4) {$e_3$};
\node at (5.6,1.5) {$e_4$};
\node at (7.6,1.5) {$e_5$};
\node at (9.9,0) {$e_6$};
\node at (8,-1.9) {$e_7$};
\node at (6.5,-1.2) {\textbf{b)}};
\end{tikzpicture}
    \caption{Bimetric tree interactions. \textbf{a)} Pairwise bimetric interaction of three vielbeins. \textbf{b)} A tree on seven vielbeins. In both cases the leaf-removal procedure solves the Lorentz constraints lapse-independently.}
    \label{fig:bimetric_trees}
\end{figure}
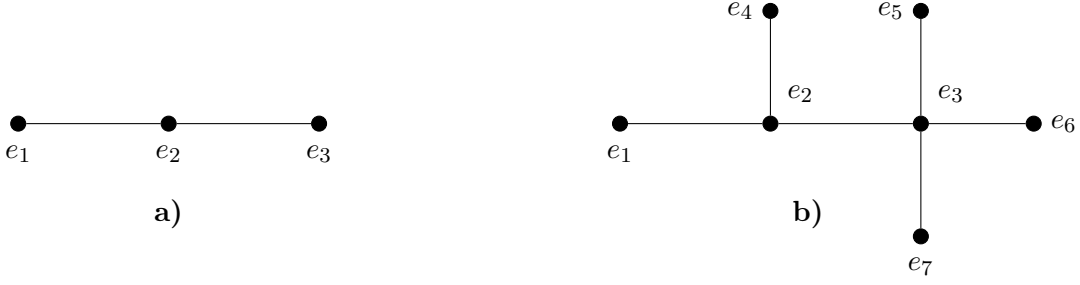

Consider $\Ncal=3$ vielbeins with the potential,
\begin{align}
\label{pairwise_bimetric}
    V = V_{\mathrm{bi}}(e_1,e_2) + V_{\mathrm{bi}}(e_2,e_3)\,,
\end{align}
where each bimetric potential is of the form \eqref{bi_pot}. This fits into both the centre and chain coupling described in \eqref{eq:pairwise} with the interaction graph figure~\ref{fig:bimetric_trees}a.

In the antisymmetrised vielbein form \eqref{bi_wedge}, the $\beta$-parameter reads,
\begin{align}
\label{chain_beta}
    \beta^{IJKL} = \beta^{IJKL}_{(12)} + \beta^{IJKL}_{(23)}\,,
\end{align}
where $\beta^{IJKL}_{(12)}$ is nonzero only for indices being 1 or 2, and $\beta^{IJKL}_{(23)}$ only for indices being 2 or 3. The individual bimetric couplings are related to $\alpha_n$ by,
\begin{align}
    \alpha^{(12)}_n = \beta^{\overbrace{ \!\scriptstyle1\ldots 1}^{4-n}\overbrace{\scriptstyle 2\ldots 2}^{n}}_{(12)}, &&
    \alpha^{(23)}_n = \beta^{\overbrace{\!\scriptstyle2\ldots2}^{4-n}\overbrace{ \scriptstyle3\ldots 3}^{n}}_{(23)}.
\end{align}
Since $\beta^{13KL}=0$ for all $K,L$, the irreducibility assumption is violated and the interaction is not subject to the rank-1 restriction. In fact, the rank can be arbitrary, but the conditions $X^a_{IJ}=0$ decouple as follows:
\begin{enumerate}[leftmargin=1.2cm]
    \item[\emph{$I=1$}:] Only $\beta^{IJKL}_{(12)}$ contributes since $\beta^{1JKL}_{(23)}=0$. So the only non-trivial $X^a_{IJ}$ is $X^a_{12}$ which implies,
    \begin{align}
        M^i_{12}=0 \qquad \Longrightarrow \qquad[E_1^{\T}\delta E_2]_{[ij]} = 0,
    \end{align}
    which determines $\Omega_1$.

    \item[\emph{$I=3$}:] Analogously, only $\beta^{IJKL}_{(23)}$ contributes and $X^a_{3J}=0$ reduces to,
    \begin{align}
        M^i_{32}=0 \qquad \Longrightarrow \qquad[E_3^{\T}\delta E_2]_{[ij]} = 0\,,
    \end{align}
    which determines $\Omega_3$.

    \item[\emph{$I=2$}:] Both sectors contribute:
    \begin{align}
        X^a_{2J} = 2\sum_{KL}\beta^{2JKL}_{(12)}\,
       M^i_{2K}\,E^a_{L\,i}
        + 2\sum_{KL}\beta^{2JKL}_{(23)}\,
        M^i_{2K}\,E^a_{L\,li}\,.
    \end{align}
    The first sum involves only $M^i_{21}=-M^i_{12}$, which vanishes by the $I=1$ solution; the second involves only $M^i_{23}=-M^i_{32}$, which vanishes by the $I=3$ solution. Therefore $X^a_{2J}=0$ is satisfied automatically, confirming the leaf-removal mechanism: $e_1$ and $e_3$ are leaves, and $e_2$ is the junction.
\end{enumerate}
In the fully covariant setting, the Lorentz constraints impose $e_1^{\T}\eta e_2 = e_2^{\T}\eta e_1$ and $e_3^{\T}\eta e_2 = e_2^{\T}\eta e_3$, which allow the definition of square-root matrices $S_{12}=\sqrt{g_1^{-1}g_2}$ and $S_{23}=\sqrt{g_2^{-1}g_3}$, giving the equivalent metric formulation $V = V_\mathrm{bi}(S_{12}) + V_\mathrm{bi}(S_{23})$.

\subsubsection{Bimetric trees}
The $\Ncal=3$ mechanism extends to arbitrary tree-shaped pairwise interactions, see Figure~\ref{fig:bimetric_trees}b. For $\Ncal$ vielbeins with potential,
\begin{align}
\label{tree_bimetric}
    V = \sum_{(IJ) \in \mathcal{T}} V_{\mathrm{bi}}(e_I,e_J)\,,
\end{align}
where $\mathcal{T}$ is a tree on $\Ncal$ vertices, the $\beta$-parameter takes the form,
\begin{align}
    \beta^{IJKL} = \sum_{( I'J') \in \mathcal{T}} \beta^{IJKL}_{(I'J')}\,,
\end{align}
with each $\beta^{IJKL}_{(I'J')}$ nonzero only when all indices are $I'$ or $J'$. The Lorentz constraints are solved by the leaf-removal procedure: each leaf $E_{I'}$ with unique neighbour $E_{J'}$ satisfies $[E_{I'}^{\T} \delta E_{J'}]_{[ij]}=0$, which determines $\Omega_{I'}$. Removing the leaf produces a smaller tree, and the procedure repeats. The identity $\sum_I X^a_{IJ}=0$ ensures that the last vielbein's equation is automatically satisfied, leaving the overall SO(3)-frame undetermined.

\subsubsection{Counterexample: bimetric cycles}
\label{sec:bi_cycle}

\begin{figure}
    \centering
\begin{tikzpicture}[scale=1.0]
\tikzstyle{every node}=[draw,shape=circle,fill=black,inner sep=2pt]
\path (0,0)   node (c1) {};
\path (2,0)   node (c2) {};
\path (1,1.5) node (c3) {};
\draw (c1) -- (c2);
\draw (c2) -- (c3);
\draw (c3) -- (c1);
\tikzstyle{every node}=[]
\node at (0,-0.4)  {$e_1$};
\node at (2,-0.4)  {$e_2$};
\node at (1, 1.9)  {$e_3$};
\end{tikzpicture}
    \caption{A 3-cycle of pairwise bimetric interactions. The absence of leaves prevents the iterative leaf-removal procedure from solving the Lorentz constraints lapse-independently.}
    \label{fig:bimetric_cycle}
\end{figure}
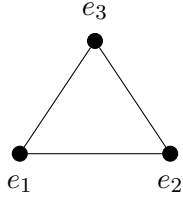

Adding a third bimetric edge between $e_1$ and $e_3$ in the $\Ncal=3$ example above, we get a cycle,\footnote{Metric cycles were studied in \cite{Nomura:2012xr, Scargill:2014wya}, and vielbein cycles in \cite{Afshar:2014dta, deRham:2015cha}.}
\begin{align}
\label{cycle_potential}
    V = V_{\mathrm{bi}}(e_1,e_2) + V_{\mathrm{bi}}(e_2,e_3) + V_{\mathrm{bi}}(e_3,e_1)\,,
\end{align}
with $\beta^{IJKL} = \beta^{IJKL}_{(12)} + \beta^{IJKL}_{(23)} + \beta^{IJKL}_{(31)}$. Every vielbein now appears in two bimetric sectors and there are no leaves, see Figure \ref{fig:bimetric_cycle}. The conditions $X^a_{IJ}=0$ read,
\begin{align}
\label{bim_loop_X}
    X^a_{1J} &= 2\sum_{KL}\beta^{1JKL}_{(12)}\,M^i_{1K}\,E^a_{L\,i} + 2\sum_{KL}\beta^{1JKL}_{(31)}\,M^i_{1K}\,E^a_{L\,i}\,, \notag\\[4pt]
    X^a_{2J} &= 2\sum_{KL}\beta^{2JKL}_{(12)}\,M^i_{2K}\,E^a_{L\,i} + 2\sum_{KL}\beta^{2JKL}_{(23)}\,M^i_{2K}\,E^a_{L\,i}\,, \notag\\[4pt]
    X^a_{3J} &= 2\sum_{KL}\beta^{3JKL}_{(23)}\,M^i_{3K}\,E^a_{L\,i} + 2\sum_{KL}\beta^{3JKL}_{(31)}\,M^i_{3K}\,E^a_{L\,i}\,.
\end{align}
The first equation determines $\Omega_1$ through a combined condition on $M^i_{12}$ and $M^i_{13}$, without setting either to zero individually. With $\Omega_1$ fixed, the second equation determines $\Omega_2$. The third equation then involves $M^i_{32}$ and $M^i_{31}$, both already fixed, constituting 6 conditions with only the identity $\sum_I X^a_{IJ}=0$ providing 3 relations. The system is overconstrained, confirming that bimetric cycles do not admit lapse-independent solutions.

Note that the vielbein potential \eqref{cycle_potential} is not equivalent to the metric potential $V_{\mathrm{bi}}(\sqrt{g_1^{-1}g_2})+V_{\mathrm{bi}}(\sqrt{g_2^{-1}g_3})+V_{\mathrm{bi}}(\sqrt{g_3^{-1}g_1})$, since the Lorentz constraints do not impose pairwise symmetrisation.

\subsection{Sums of determinant interactions}

\subsubsection{Two determinants sharing one vielbein}

\begin{figure}
    \centering
\begin{tikzpicture}[scale=1.0]
\tikzset{
    vnode/.style={draw, circle, fill=black, inner sep=2pt},
    inode/.style={draw, circle, fill=white, inner sep=3pt, minimum size=12pt},
}
\node[vnode] (c) at (0,0) {};
\node[inode] (Lcirc) at (-3,0) {};
\node[inode] (Rcirc) at (3,0) {};
\draw (c) -- (Lcirc);
\draw (c) -- (Rcirc);
\def\r{1.85}
\foreach \angle/\name in {
40/l1, 86.667/l2, 133.333/l3, 180/l4, 226.667/l5, 273.333/l6, 320/l7} {
    \node[vnode] (\name) at ($(-3,0)+(\angle:\r)$) {};
    \draw (Lcirc) -- (\name);
}
\foreach \angle/\name in {
140/r1, 93.333/r2, 46.667/r3, 0/r4, 313.333/r5, 266.667/r6, 220/r7} {
    \node[vnode] (\name) at ($(3,0)+(\angle:\r)$) {};
    \draw (Rcirc) -- (\name);
}
\node at (0, -0.42) {$e_1$};
\end{tikzpicture}
    \caption{Two determinant sectors sharing the vielbein $e_1$. All other vielbeins are leaves.}
    \label{fig:two_det}
\end{figure}
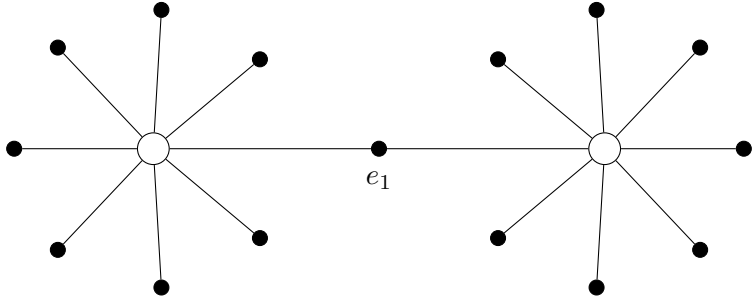

Consider two groups of vielbeins determined by the index sets $\Acal_1$ and $\Acal_2$ sharing only $e_1$, as in Figure~\ref{fig:two_det}, and with the potential,
\begin{align}
    V = 2m^4\bigg[\det\Big(\sum_{I\in \Acal_1}\beta^I_1 e_I\Big) + \det\Big(\sum_{I\in \Acal_2}\beta^I_2 e_I\Big)\bigg],
\end{align}
where $\beta^I_1\neq 0$ only for $I\in \Acal_1$ and $\beta^I_2\neq 0$ only for $I\in \Acal_2$. The interaction tensor $\beta^{IJKL} = \beta^I_1\beta^J_1\beta^K_1\beta^L_1 +\beta^I_2\beta^J_2\beta^K_2\beta^L_2$ has rank $R=2$ and is reducible. Note that all vielbeins but $e_1$ are leaves.

With $U_1=\sum_{K\in \Acal_1}\beta^K_1 E_K$ and $U_2=\sum_{K\in \Acal_2}\beta^K_2 E_K$, the conditions $X^a_{IJ}=\beta^J_1 t^a_{1I}+\beta^J_2 t^a_{2I}=0$ decouple:
\begin{enumerate}[leftmargin=2.2cm]
    \item[\emph{$I\in \Acal_1\setminus\{1\}$}:] Since $\beta^I_2=0$, only $t^a_{1I}=0$ contributes, giving,
    \begin{align}
    \label{A_set_sol}
        [E_I^{\T}\delta\, U_1]_{[ij]}=0,
    \end{align}
    which determines $\Omega_I$ for $I\in \Acal_1\setminus \{1\}$.

    \item[\emph{$I\in \Acal_2\setminus\{1\}$}:] Since $\beta^I_1=0$, only $t^a_{2I}=0$ contributes, giving,
    \begin{align}
    \label{B_set_sol}
        [E_I^{\T}\delta\, U_2]_{[ij]}=0,
    \end{align}
    which determines $\Omega_I$ for $I \in \Acal_2\setminus\{1\}$.

    \item[\emph{$I=1$}:] Both terms contribute. However, the identity $\sum_{I\in \Acal} \beta^I [E_I^{\T}\delta U_1]_{[ij]}=0$ together with \eqref{A_set_sol} forces $[E_1^{\T}\delta U_1]_{[ij]}=0$. Similarly, \eqref{B_set_sol} forces $[E_1^{\T}\delta U_2]_{[ij]}=0$. Therefore $X^a_{1J}=0$ is satisfied automatically.
\end{enumerate}
The two sectors decouple completely, each solved independently by the leaf-removal procedure.

\subsubsection{Trees of determinant couplings}
The above construction generalises to $R$ groups $\Acal_1,\ldots,\Acal_R$, each carrying a determinant coupling $V_\mathrm{det}(u_r)$ with $u_r=\sum_{I\in \Acal_r}\beta^I_r e_I$, provided any two groups share at most one vielbein and the interaction graph $\Gcal$ is a tree. The interaction tensor is $\beta^{IJKL} = \sum_{r} \beta^I_r\beta^J_r\beta^K_r\beta^L_r$, which is reducible since $\beta^{IJKL}=0$ whenever $I$ and $J$ belong to different sectors without sharing a vielbein.

The leaf-removal procedure applies directly: each leaf vielbein $e_{I'}$ in sector $\Acal_r$ yields only the condition $[E_{I'}^{\T}\delta U_r]_{[ij]}=0$. Once all leaves of a sector are removed, the identity $\sum_{I\in \Acal_r} t^a_{rI}=0$ ensures the shared vielbein's contribution vanishes, and $\Acal_r$ can be removed from $\Gcal$. The procedure repeats until one vielbein remains, whose equation is trivially satisfied by $\sum_I X^a_{IJ}=0$.

\subsubsection{Determinant plus bimetric interaction}

Consider $\Ncal=4$ vielbeins with potential (Figure~\ref{fig:convention_and_det_bi}b),
\begin{align}
\label{det_plus_bi}
    V = 2m^4\det\big(\beta^1 e_1 + \beta^2 e_2 + \beta^3 e_3\big) + V_{\mathrm{bi}}(e_3,e_4)\,,
\end{align}
where $e_3$ is shared between the determinant sector $\Acal=\{1,2,3\}$ and the bimetric edge $(3,4)$. The interaction tensor is $\beta^{IJKL} = \beta^I\beta^J\beta^K\beta^L + \beta^{IJKL}_{(34)}$, which is reducible if $\beta^4=0$ and $\beta^{IJKL}_{(34)}=0$ if $IJKL\in\{1,2\}$. With $U=\beta^1 E_1+\beta^2 E_2+\beta^3 E_3$, the conditions decouple:
\begin{enumerate}[leftmargin=2.2cm]
    \item[\emph{$I=1,2$}:] Only the determinant term contributes, giving $[E_I^{\T}\delta U]_{[ij]}=0$, which determines $\Omega_1$ and $\Omega_2$.

    \item[\emph{$I=4$}:] Only $\beta^{IJKL}_{(34)}$ contributes, giving $[E_4^{\T}\delta E_3]_{[ij]}=0$, which determines $\Omega_4$.

    \item[\emph{$I=3$}:] Both terms contribute, so that,
    \begin{align}
        X^a_{3J}= \sum_{KL}\beta^3\beta^J\epsilon^{ijk}[E_3^{\T}\delta U]_{ij}U^a_{~k} + 2\sum_{KL}\beta^{3JKL}_{(34)}\,M^i_{3K}\,E^a_{L\,i},
    \end{align}
    but the identity  $\sum_I \beta^I[E_I^{\T}\delta U]_{[ij]}$ together with the solutions for $I=1,2$ implies $[E_3^{\T}\delta U]_{[ij]}=0$. Similarly,  $M^i_{34}=-M^i_{43}=0$ by the $I=4$ solution. Therefore $X^a_{3J}=0$ is automatically satisfied.
\end{enumerate}
This confirms the mixed leaf-removal: $e_1$, $e_2$, and $e_4$ are leaves, and the junction $e_3$ is automatically satisfied.

\section{Determinant coupling as a sum of higher rank interactions}
\label{app:curing}

In this appendix we illustrate how the determinant interaction \eqref{det_coupling}, when expanded in terms of antisymmetrised wedge products, decomposes into a sum of pieces that are individually inconsistent, but which combine into a ghost-free interaction thanks to the specific coefficients dictated by $\beta^{IJKL}=\beta^I\beta^J\beta^K\beta^L$. For pedagogical clarity we work in $2{+}1$ dimensions with $\Ncal=3$ vielbeins, where the expansion contains only a few terms. The same mechanism operates in $3{+}1$ dimensions, which we comment on at the end.

\subsection{Expanding the determinant}

In $D=3$ spacetime dimensions, the Hinterbichler--Rosen interaction reads,
\begin{align}
    V = \frac{2m^3}{3!}\sum_{IJK=1}^{\Ncal}\beta^{IJK}\epsilon_{ABC}\epsilon^{\alpha \beta \gamma}e^A_{I\, \alpha}e^B_{J\, \beta}e^C_{K\, \gamma},
\end{align}
with totally symmetric couplings $\beta^{IJK}$. For $\Ncal=3$ and the determinant choice $\beta^{IJK}=\beta^I\beta^J\beta^K$, the potential becomes,
\begin{align}
    V_{\mathrm{det}} = 2m^3\det\big(\beta^1 e_1 + \beta^2 e_2 + \beta^3 e_3\big).
\end{align}
Expanding the determinant and grouping terms by which vielbeins appear, we find,
\begin{align}
\label{det_expansion}
    V_{\mathrm{det}} = 2m^3\Big[\, V_{\mathrm{CC}} + V^{(12)}_{\mathrm{bi}} + V^{(23)}_{\mathrm{bi}} + V^{(31)}_{\mathrm{bi}} + V^{(123)}\,\Big],
\end{align}
where $V_{\mathrm{CC}}$ collects the single-vielbein terms $(\beta^I)^3\det e_I$ which contribute only cosmological constants for each $g_I$ and play no role in the Lorentz constraints. The pairwise pieces are,
\begin{align}
    V^{(IJ)}_{\mathrm{bi}} = \tfrac{3}{2!}\beta^I \beta^J\epsilon_{ABC}\epsilon^{\alpha \beta \gamma}\Big[\beta^I e^A_{I\, \alpha}e^B_{I\, \beta}e^C_{J\, \gamma} + \beta^J e^A_{I\, \alpha}e^B_{J\, \beta}e^C_{J\, \gamma}\Big],
\end{align}
and the genuinely three-field antisymmetrised piece is,
\begin{align}
    V^{(123)} = 6\beta^1\beta^2\beta^3\epsilon_{ABC}\epsilon^{\alpha \beta \gamma}e^A_{1\, \alpha}e^B_{2\, \beta}e^C_{3\, \gamma}.
\end{align}
Now, each of the non-cosmological pieces in \eqref{det_expansion}, taken on its own, corresponds to an interaction that is known to be inconsistent:
\begin{itemize}
    \item The three bimetric pieces $V^{(12)}_{\mathrm{bi}}$, $V^{(23)}_{\mathrm{bi}}$, $V^{(31)}_{\mathrm{bi}}$, together but without the wedge term, form a pairwise bimetric $3$-cycle. This is the $D=3$ analogue of Appendix~\ref{sec:bi_cycle}, and its Lorentz constraints do not admit lapse-independent solutions.
    \item The fully antisymmetrised term $V^{(123)}$ is the $D=3$ analogue of the $\beta^{1234}$-interaction of Appendix~\ref{beta_1234}: a single wedge involving all vielbeins whose interaction graph has no leaves. By the same argument as in Appendix~\ref{beta_1234}, it does not admit lapse-independent solutions on its own.
\end{itemize}

\subsection{Cancellation of obstructions}

Despite each ingredient in \eqref{det_expansion} being individually inconsistent, their sum is precisely the determinant interaction $V_{\mathrm{det}}$, which by the analysis of Section~\ref{sec:restric} has $\beta^{IJK}$ of symmetric rank $R=1$ and therefore admits lapse-independent Lorentz constraints \eqref{eetau=}.

The cancellation of obstructions can then be understood as follows. The bimetric $3$-cycle, in isolation, produces conditions of the form \eqref{bim_loop_X},
\begin{align}
    X_{1J}&\sim c_{12}M_{12} + c_{13}M_{13},\notag \\
    X_{2J}&\sim c_{21}M_{21} + c_{23}M_{23},\\
    X_{3J}&\sim c_{31}M_{31} + c_{32}M_{32},\notag
\end{align}
with $M_{IJ}$ the $D=3$ analogue of \eqref{M_def} and coefficients $c_{IJ}$ built from pairs $\beta^I,\beta^J$. Each equation mixes two distinct $M_{IJ}$ with coefficients that do not combine into a common composite vielbein, and the system is overconstrained. The wedge piece $V^{(123)}$ contributes additional cross-terms proportional to $\beta^1\beta^2\beta^3$. With the determinant coefficients $\beta^{IJK}=\beta^I\beta^J\beta^K$, these are precisely the missing contributions needed to complete each $X_{IJ}$ into the single rank-$1$ combination
\begin{align}
    X_{IJ}\propto \beta^I\beta^J\big[E_I^{\T}\delta\, U\big]_{[ij]},\qquad U = \sum_K\beta^K E_K,
\end{align}
which is the $D=3$ specialisation of \eqref{C=Nbetat} for $R=1$. The obstruction of the bimetric cycle and that of the wedge interaction thus cure each other.

\subsection{Generalisation to \texorpdfstring{$3{+}1$}{3+1} dimensions}

The same pattern occurs in $D=4$. For $\Ncal=4$, expanding $V_{\mathrm{det}} = 2m^4\det(\sum_I \beta^I e_I)$ yields, besides cosmological constant terms $(\beta^I)^4\det e_I$:
\begin{itemize}
    \item six pairwise bimetric-type interactions $V^{(IJ)}_{\mathrm{bi}}$ for $I<J$, which together form a bimetric interaction graph containing multiple cycles and hence is inconsistent on its own;
    \item four $3$-field interactions, each involving three distinct vielbeins, corresponding to reductions of the $\beta^{IJKL}$-interaction with one repeated index, inconsistent by the argument of Appendix~\ref{beta_1234};
    \item one fully antisymmetrised $4$-wedge $\beta^1\beta^2\beta^3\beta^4\epsilon_{ABCD}\epsilon^{\alpha \beta \gamma \delta}e^A_{1\, \alpha}e^B_{2\, \beta}e^C_{3\, \gamma}e^D_{4\, \delta}$, which is exactly the $\beta^{1234}$-interaction of Appendix~\ref{beta_1234}.
\end{itemize}
Each of these pieces is individually inconsistent, yet assembled with the coefficients $\beta^{IJKL}=\beta^I\beta^J\beta^K\beta^L$ they combine into the consistent determinant interaction. As shown in Section~\ref{sec:restric}, this cancellation is only possible because symmetric rank $R=1$ is precisely the condition under which the obstructions of the individual sectors cancel against one another.

\printbibliography

\end{document}